\documentclass[showpacs,aps,floatfix]{revtex4}

\usepackage[final]{graphicx}
\usepackage{t1enc}
\usepackage{bm}
\begin{document}

\title{Two dimensional Ising model with long-range competing interactions}

\date{\today}

\author{Sergio A. Cannas}
\email{cannas@famaf.unc.edu.ar}
\affiliation{Facultad de  Matem\'atica, Astronom\'{\i}a  y F\'{\i}sica, Universidad
Nacional de C\'ordoba, \\ Ciudad Universitaria, 5000 C\'ordoba, Argentina}
\altaffiliation{Member of CONICET, Argentina}
\author{Pablo M. Gleiser}
\email{gleiser@tero.fis.uncor.edu}
\affiliation{Facultad de  Matem\'atica, Astronom\'{\i}a  y F\'{\i}sica, Universidad
Nacional de C\'ordoba, \\ Ciudad Universitaria, 5000 C\'ordoba, Argentina}
\altaffiliation{Member of CONICET, Argentina}
\author{Francisco A. Tamarit}
\email{tamarit@famaf.unc.edu.ar}
\affiliation{Facultad de  Matem\'atica, Astronom\'{\i}a  y F\'{\i}sica,
Universidad Nacional de C\'ordoba, \\ Ciudad Universitaria,
5000 C\'ordoba, Argentina}
\altaffiliation{Member of CONICET, Argentina}


\begin{abstract}
The two-dimensional Ising model with competing short range
ferromagnetic interactions and long range antiferromagnetic
interactions is perhaps the most simple one containing the minimal
microscopic ingredients necessary for an appropriate description
of the macroscopic properties of ultrathin films and
quasi--two--dimensional magnetic materials. Despite such relative
simplicity, the frustration introduced by the competition between
interactions generates complex behaviors that have eluded, up to now,
a complete understanding of its general properties.
In this work we review recent advances in the understanding of
both equilibrium and non-equilibrium properties of the model. This
includes a detailed description of several known properties of the
thermodynamical phase diagram, as well as the existence of several
types of metastable states and their influence in the low
temperature dynamics.
\end{abstract}


\maketitle

\section{Introduction}
The Ising model with arbitrary interactions is perhaps the most
successful one in statistical physics. Despite its (relative)
simplicity, it has been used to model such a variety of complex
systems exhibiting cooperative phenomena that it can be regarded
with fairness  as a paradigm of a model system. Besides its
multiple applications in theoretical physics (in fact, Onsager's
solution of the two-dimensional model was the cornerstone of the
modern theory of critical phenomena) one of its most conspicuous
application in condensed matter (and its original motivation) is
related to the description of anisotropic magnetic materials.
While mostly applied in the past decades to bulk magnetic
materials, the advances in film growth techniques, such as atomic
or molecular epitaxy, has raised a renewed interest in the
behavior of the two-dimensional version of the model. In
particular, the study of thin magnetic films and quasi
two--dimensional magnetic materials (like the rare earth layers
that occur in perovskite structures of RE${\rm Ba}_2$${\rm
Cu}_3$${\rm O}_{7-\delta}$, RE being a rare earth of the
lanthanide series \cite{DeBell})  has experimented an increasing
interest during the last years. Part of this interest is obviously
motivated by the great amount of applications they find nowadays
in many different technological fields, such as data storage, catalysis
and electronic uses are only a few examples, among many others.
Nevertheless, they have also caught the attention of both
theoretical and experimental physicists due to the rich
insight they provide into the fundamental role that
microscopic interactions play in determining the macroscopic
properties of a material. One of the most interesting applications
of the two-dimensional Ising model is related to the theoretical
description of ultra-thin magnetic films, like metal films on
metal substrates (e.g. Fe on Cu \cite{Pappas}, Co on Au
\cite{Allenspach}, see also \cite{DeBell} for a recent review
on the topic). If the magnetic film is thin enough (less than
approximately five monolayers) the atomic magnetic moments tend to
align out of the plane defined by the proper film. This occurs
because the surface anisotropy overcomes the anisotropy of the
dipolar interactions, which favor in--plane ordering. Under these
circumstances, one can then model the local magnetic momenta of
the material by using Ising variables.

Any realistic theoretical description of a magnetic thin film must
include, besides the usual short--range exchange interactions,
also the long--range antiferromagnetic dipolar interactions. It is
worth mentioning that dipolar interactions have been usually
neglected, specially in thermodynamical studies, due to the small
magnitude of dipolar interactions relative to the magnitude of the
exchange interactions. Nevertheless, it is a well established fact
that dipolar interactions can give place to very rich
phenomenological scenarios, concerning both thermodynamical
and nonequilibrium properties. In particular, when both
interactions are present, the system is inherently frustrated, and
many of the interesting static and dynamical properties of systems
with dipolar interactions result, precisely, from this property.

In this paper we will review the recent advances in the study of
equilibrium ({\it i.e., thermodynamical}) and out of equilibrium
properties of the Ising model with competition between
short--range ferromagnetic exchange interactions and long--range
antiferromagnetic dipolar interactions defined on a square lattice
and ruled by the following dimensionless Hamiltonian:
\begin{equation}
{\cal H}= - \delta \sum_{<i,j>} \sigma_i \sigma_j + \sum_{(i,j)}
\frac{\sigma_i \sigma_j}{r^3_{ij}} \label{Hamilton1}
\end{equation}
where $\sigma=\pm 1$ and $\delta$ stands for the quotient between
the exchange $J_0$ and the dipolar $J_d$ interactions parameters,
i.e., $\delta = J_0/J_d$. The first sum runs over all pairs of
nearest neighbor spins and the second one over all distinct pairs
of spins of the lattice. $r_{ij}$ is the distance, measured in
crystal units, between sites $i$ and $j$. The energy is measured
in units of $J_d$, which is assumed to be always positive
(antiferromagnetic); hence $\delta >0$ means ferromagnetic
exchange couplings.

\section{Thermodynamical properties}

\subsection{Ground state}

The basic features of the finite temperature phase diagram
associated with Hamiltonian (\ref{Hamilton1}) were first
derived by MacIsaac and coauthors in 1995 \cite{Macisaac}. In that
work the existence of equilibrium striped states was firmly
established for the first time. They showed that the ground state
of Hamiltonian (\ref{Hamilton1}) is indeed the striped one,
provided that the relative strength of the interactions $\delta$
exceeds some positive value $\delta_a \approx 0.425$
\cite{differences}. If $\delta<\delta_a$ the ground state is the
antiferromagnetic one. If
 $\delta>\delta_a$ the ground state is composed by antialigned stripes of width $h$, which
depends on $\delta$. MacIsaac and coworkers showed that for every
finite value of $\delta>\delta_a$  there exists always a value of
$h$ such that the corresponding striped state has less energy than
both the ferromagnetic and the checkerboard states, which had been
previously proposed as the ground state for this
system \cite{Czech}. Moreover, they showed that this is true for
{\it any} value of $\delta$. In other words, the presence of the
dipolar interactions, even with an infinitesimal strength,
suppress the ferromagnetic ordering, favoring the formation of
stripe domains. The also showed that for large values
 of $\delta$ the equilibrium width increases exponentially $h \sim e^{\delta/2}$.
Although not rigorously demonstrated up to now, a very large
amount of numerical evidence supports the results of MacIsaac and
coauthors.

\subsection{Monte Carlo finite temperature simulations}
\label{simulaciones}

Most of the knowledge about the finite temperature equilibrium
properties of the model has
 been obtained by means of Monte Carlo numerical simulations on finite lattices. These
have
been performed  on square lattices of $N=L\times L$
sites \cite{Macisaac,Booth,Gleiser1,Cannas1} using Metropolis or heat bath algorithms.
Monte carlo simulations has been vastly used to study finite temperature properties of
magnetic lattice spin models.  However, performing Monte Carlo simulations in
system (\ref{Hamilton1}) involves some particular subtleties  to take
into account, due to the
presence of dipolar interactions.

First of all, the modulated nature of the ground state implies that the linear size $L$ of
 the system must be chosen commensurate with the period of the modulation associated with
the particular value of $\delta$ under consideration; otherwise,
an artificial frustration is introduced.

A second point concernes the structure of the ground state
as the value of $\delta$ is increased. For relatively large values
of $\delta$,  striped states of widths similar to that of the
ground state have very low energies. This may generate (as we will
see later) multiple metastable states at low temperature,  making
it very difficult to equilibrate and therefore to determine the
truly thermodynamically stable state.

Other complications arise as direct consequences of the long-range character of the
dipolar interactions. Since every spin in the lattice interacts with each other, flipping
a spin does not affect only a few neighbor spins, but {\it all} the rest of the spins in
the lattice. This means that most of the powerful algorithms developed in the last years
to improve the efficiency of Monte Carlo simulations (like block algorithms) do not apply,
since most of them rely on the finite range of the interactions. To further complicate the
 situation, the antiferromagnetic nature of the dipolar term does not allow the
application
of some specially designed  algorithm for systems with long-range
ferromagnetic interaction \cite{Luijten}. This means that, except from some recent works
on  the triangular lattice\cite{Stoycheva}, most of
the simulations of this model are based on simple Metropolis  or heat bath
algorithms, in which the typical number of operations to perform a
Monte carlo step  goes as $N^2$ rather than $N$, as in systems
with short range interactions.

Other consequences of the long-range character of the dipolar
interactions are the strong finite-size effects, since in this
case every spin feels {\it directly} the influence of the border.
As in any type of system, the best way to diminish border effects
is to use periodic boundary conditions.  While both the
interpretation and implementation of periodic boundary conditions
are straightforward in systems with short range microscopic
interactions, their usage in systems with long-range interactions
is more subtle. In a general situation, there are two alternative
representations of a finite two-dimensional system with periodic
boundary conditions.  On one hand, we can visualize the system as
a torus, where a given spin interacts with its closer neighbors
defined by the topology of the torus, up  to a certain range of
interaction, that is, to consider a finite translationally
invariant system. On the other hand, we can think that we have an
infinite system, where  the original finite system has been
replicated infinite times in all the coordinate directions. In
other words, we can think that the infinite lattice has been
partitioned into cells of size $N=L^2$ and that we have chosen
only the periodic solutions of the problem, with periodicity $L$
in all the coordinate directions. While the difference between
both views is just a matter of interpretation  for systems with
short-range interactions, the situation changes for long-range
interacting systems. According to the second scheme, a given spin
will interact with {\em infinite replicas of everyone of the rest
of the spins} and we can express the effective interaction between
two spins inside the system as an infinite sum over replicas (this
includes a ``self-interaction'' term, that accounts for the
interaction of every spin with its own replicas). On the other
hand, in the closed topology of the first scheme it is consistent
to consider that every pair of spins inside the system interacts
only through its {\it minimal} distance over the torus. This
corresponds to the minimal truncation of the infinite series of
the previous scheme and it is sometimes called the {\it first
image convention}.  While the efficiency of both type of
schemes can be very different in systems that presents low
temperature ferromagnetic ordering, in the  case of Hamiltonian
(\ref{Hamilton1}) the differences are minimal \cite{Cannas2}, with
a little improvement obtained with the first scheme. The usage of
a particular scheme is then a matter of convenience. However,
while the usage of the second scheme is straightforward, the
implementation of the first is more involved, since the infinite
series for the effective interactions are slowly convergent. The
usual way of handling them numerically  is to adapt the Ewald sums
technique \cite{DeBell,Kretschmer}, originally derived for systems
of interacting charged particles \cite{Allen,Frenkel}.

\subsection{Low temperature equilibrium properties: metastable striped states}

The finite temperature phase diagram in the $(\delta,T)$ space was
first calculated using Monte Carlo simulations on small lattices
($L=16$) by MacIsaac and coauthors in \cite{Macisaac}. Further
improvements at low temperature were obtained by Gleiser and
coauthors in \cite{Gleiser1}. The numerical simulations showed
that both the antiferromagnetic and the different striped states
remain stable at finite temperatures below some $\delta$-dependent
critical temperature $T_c(\delta)$. At $T=T_c(\delta)$ the
specific heat

\begin{equation}
C_L = \frac{1}{NT^2} \left( \left< H^2 \right> -\left< H \right>^2\right)
\label{calor1}
\end{equation}

\noindent presents a peak, indicating that the system undergoes a
phase transition into a disordered phase. Both the nature of the
disordered fase, as well as the order of the phase transition
present  several subtleties and will be considered in section
\ref{hightemperature}. We will consider now the properties of the
system at temperatures well below the critical one. We will
concentrate our attention in the low $\delta$ region of the phase
diagram, that is,  where the first striped states emerge after the
antiferromagnetic one, when the value of $\delta$ crosses
$\delta_a=0.425$. The transition points between the different
striped states at $T=0$ can be estimated by a numerical
calculation of the striped states energies as a function of
$\delta$ in finite size lattices of increasing linear sizes $L$
\cite{Macisaac}.
We will denote by $hi$ the striped state of width $h=i$, with
$i=1,2\ldots$. The state hi in a translationally invariant lattice
has a degeneracy $4i$, due to parallel translations along a
coordinate axis and to $\pi/2$ rotations of the stripe pattern.
Let $\delta_{i,i+1}$ be the transition point between the striped
states hi and ${\rm hi+1}$. We have for the first striped states
that $\delta_{1,2}\approx 1.26$, $\delta_{2,3}\approx 2.2$ and
$\delta_{3,4}\approx 2.8$.

We now consider the transition between the striped states h1 and
h2 at finite temperatures \cite{Gleiser1}. The different striped
phases can be characterized by the order parameter:
\begin{equation}
m_h \equiv \frac{1}{N} \sum_{x=1}^N \sum_{y=1}^N \left( -1 \right)^{f_h(x)} \sigma_{xy}
\label{orderp1}
\end{equation}
\noindent where the function $f_h(x)\equiv (x-mod(x,h))/h$ takes
odd and even values with periodicity $h$ \cite{DeBell}; $m_h$
measures the overlap of the spin configuration with a vertical
stripe pattern of width $h$. The thermodynamical averages
$M_h=\left< m_h \right>$ and its associated susceptibilities
\begin{equation}
 \chi_h \equiv \left< m_h^2 \right> - M_h^2
\label{suscep1}
\end{equation}
\noindent can be calculated through Monte Carlo simulations.
\begin{figure}
\begin{center}
\includegraphics[width=9cm,height=8cm,angle=0]{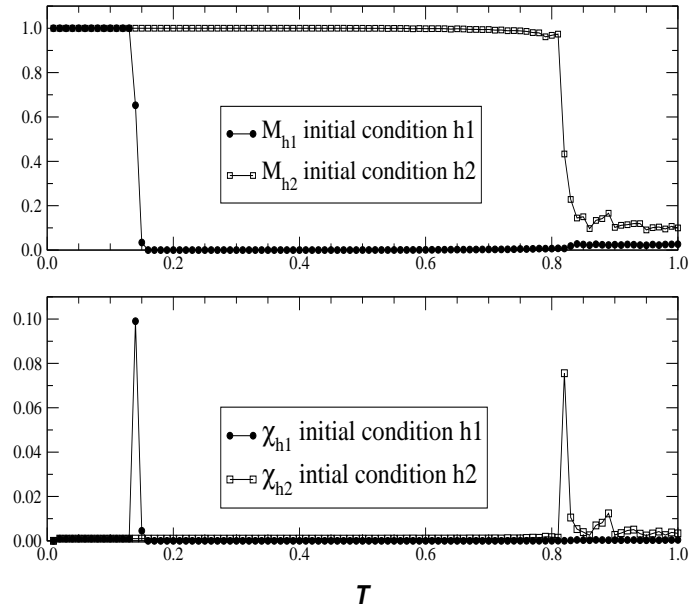}
\caption{\label{stability1} Order parameters (\ref{orderp1}) and associated
susceptibilities (\ref{suscep1})  {\it vs.} $T$ when the system is heated starting at
$T=0$ from an initial (vertical)  stripe configuration  for $L=24$ and $\delta=2$; open
squares: $M_{h2}$ for an initial configuration h2; full circles: $M_{h1}$ for an initial
configuration h1.}
\end{center}
\end{figure}
\begin{figure}
\begin{center}
\includegraphics[width=9cm,height=6cm,angle=0]{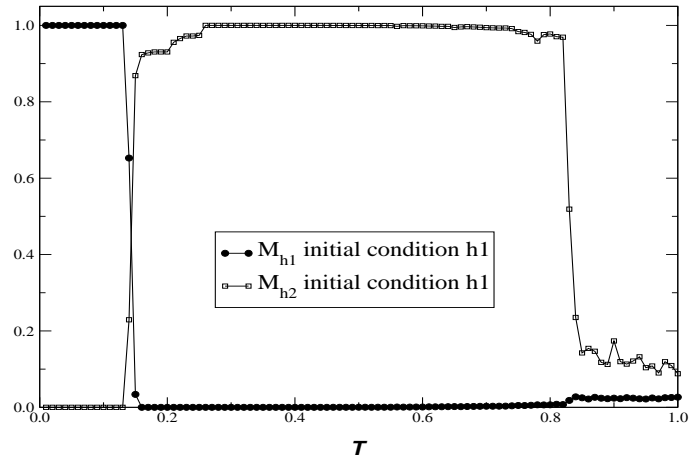}
\caption{\label{stability2} $M_{h1}$ and $M_{h2}$ when the system is heated starting at
$T=0$ from an initial  (vertical)  stripe configuration  for $L=24$ and $\delta=2$.}
\end{center}
\end{figure}
The stability of the striped phases can be analyzed by heating the
system from zero temperature, that is, by performing equilibrium
Monte Carlo simulations in a sequence of increasing temperature
values starting from $T=0$, where the initial spin configuration
at
 every temperature is taken as the last configuration of the previous one. The simulations
start at $T=0$ from an ordered configuration. In
Fig.\ref{stability1} we show an example for $\delta=2$,
corresponding to an h2 ground state. In this case  two simulations
were performed in parallel, one starting from the h1 and the other
from the h2
 (vertical) striped states, and in each case   $M_{h1}$ and $M_{h2}$ were respectively
measured, as well as the corresponding susceptibilities. In
Fig.\ref{stability2} we show a simultaneous measurement of
$M_{h1}$ and $M_{h2}$ in a simulation started from the h1 state.
When starting from h2 the system remains in this state up to the
critical temperature (i.e., that of the peak in the specific heat)
where $M_{h2}$ suddenly drops to a small value, with an associated
peak in the corresponding susceptibility. The same behavior is
reversely obtained by slowly cooling the system from high
temperatures. This
 shows that the h2 state is thermodynamically stable. On the other hand, when starting the
simulation from the h1 state it remains stable up to certain
temperature below the critical one, where it destabilizes and the
h2 pattern emerges (see Fig.\ref{stability2}), showing the
metastable nature of the h1 phase for this particular value of
$\delta$. The reverse behavior is observed in the region
$\delta<\delta_{1,2}$, where the h2 phase appears metastable down
the certain value $\delta \sim 0.7$. Repeating this procedure
allows the calculation \cite{Gleiser1} of the stability lines for the
h1 and h2 phases as a function of $\delta$. The results are shown
in the phase diagram of Fig.\ref{phased}.
\begin{figure}
\begin{center}
\includegraphics[width=11cm,height=8cm,angle=0]{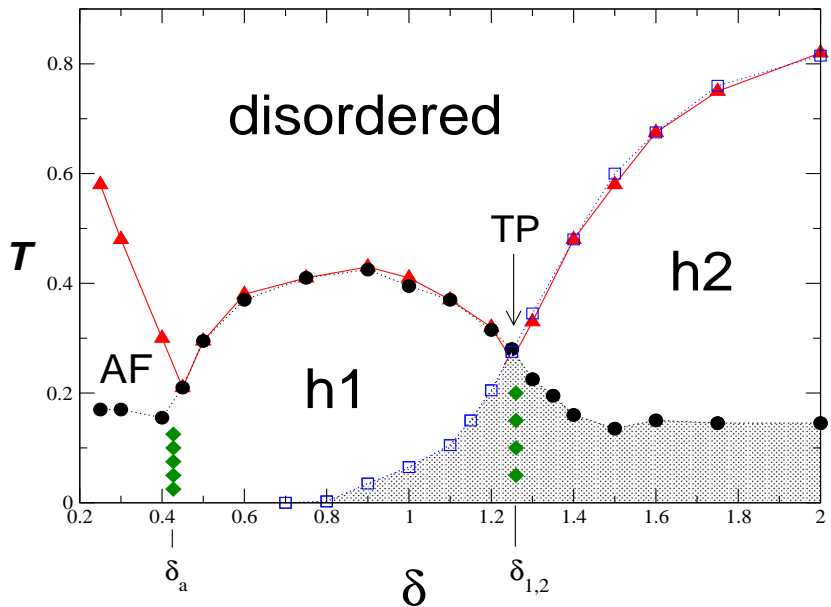}
\caption{\label{phased} $(T,\delta)$ phase diagram for low values
of $\delta$ obtained from Monte Carlo simulations in $L=24$
lattices. Triangles: critical temperature obtained from the
maximum in the specific heat; circles: stability line of the h1
striped phase; open squares:  stability line of the h2 striped
phase; diamonds: first order transition lines between low
temperature ordered phases. TP indicates a triple point.}
\end{center}
\end{figure}
\begin{figure}
\begin{center}
\includegraphics[width=8cm,height=6cm,angle=0]{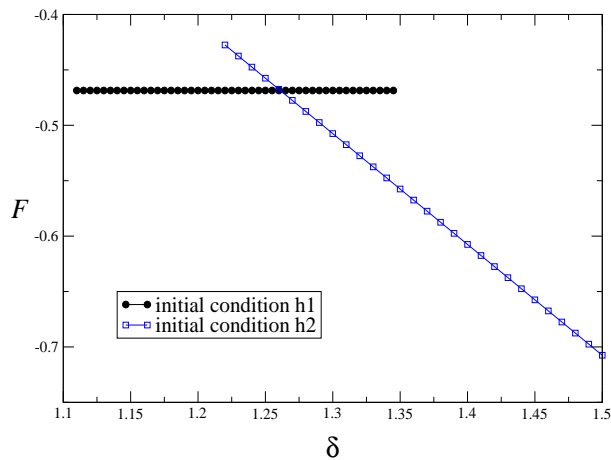}
\caption{\label{fenergy} Free energy per spin of states h1 and h2 {\it vs.} $\delta$ for
$T=0.2$ and $L=24$.}
\end{center}
\end{figure}
The coexistence of different states indicates that the transition
between both ordered phases at low temperature is of the first
order type. The first order transition line can be located by
calculating the free energy per spin of each phase, which can be
obtained by numerical integration

\begin{equation}
F(T,\delta) = U - T \int_0^T \frac{C(T')}{T'} dT'
\end{equation}

\noindent where $U\equiv \left<H\right>/N$ and  $C(T)$ is obtained by heating from $T=0$
up to the reference temperature $T$ for every value of $\delta$.  For a given value of
$T$, $F(T,\delta)$ is computed for different values of $\delta$ in each
phase. The transition point is calculated as the crossing point of the two curves, as
depicted in the example presente in  Fig. \ref{fenergy}. The first
order transition line
between h1 and h2 phases is shown in diamonds in Fig.\ref{phased}. Following the same
procedure another first order vertical line is encountered between the antiferromagnetic
and the h1 phases \cite{Gleiser2}. Other vertical transition lines appear between ordered
phases of higher $h$ values \cite{Macisaac}.

\begin{figure}
\begin{center}
\includegraphics[width=9cm,height=7cm,angle=0]{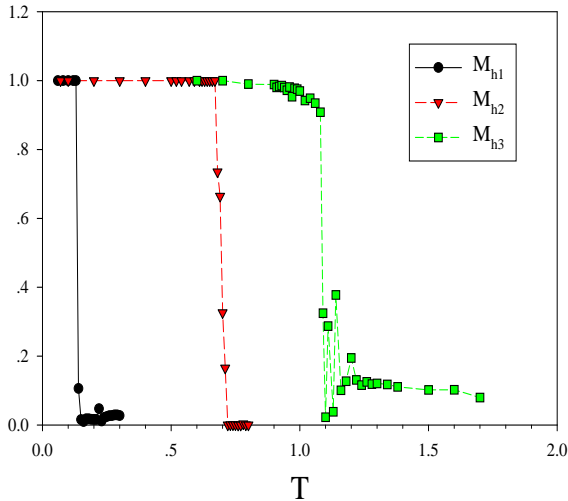}
\caption{\label{stability3} Order parameters (\ref{orderp1}) {\it
vs.} $T$ when the system is heated starting at $T=0$ from an
initial (vertical) stripe configuration  for $L=24$ and
$\delta=2.6$; circles: $M_{h1}$ for an initial configuration h1;
triangle: $M_{h2}$ for an initial configuration h2; squares:
$M_{h3}$ for an initial configuration h3.}
\end{center}
\end{figure}

We now turn our attention to the general features of the phase
diagram shown in  Fig. \ref{phased}. We see that the first order line
between phases h1 and h2 joins the point where the two stability
lines cross, marked as TP in Fig. \ref{phased}. Above the point TP
both stability lines coincide with the maxima in the specific heat
and thus with the transition lines to the disordered phase. As we
will see in \ref{hightemperature}, those transitions are also of
the first order type and, therefore, TP is actually a {\it triple
point}. Below the triple point we have a metastability region,
which appears shadowed in Fig.\ref{phased}. The corresponding
spinodal lines are given by the continuation of the order-disorder
transition lines below TP. Further evidence of the metastable
nature of h1 and h2 phases in that region can be obtained be
analyzing the relaxation of mixed states in the shadow  region of
the phase diagram \cite{Gleiser1}. Following the same procedures
described above, the coexistence of striped
states with larger values of $h$ at higher values of $\delta$ can be
verified.
Notice that the spinodal line of h1 in the h2 region decays very
slowly as $\delta$ increases; actually, it continues almost
horizontal for a wide range of values of $\delta$. Below this line
the coexistence of multiple striped states can be found, for
instance, between h1, h2 and h3 when $\delta_{2,3}< \delta <
\delta_{3,4}$, as can be appreciated in the example for $\delta=2.6$
shown in Fig.\ref{stability3}. Coexistence of a larger number of
striped states can be expected for higher values of $\delta$.
Moreover, other type of stable or metastable configurations (like
mixtures of stripes of different widths) can also be expected in
that region, as observed in a related three-dimensional model with
competing nearest neighbors ferromagnetic interactions and
long-range antiferromagnetic Coulomb-like
interactions \cite{Grousson}. As we will se in \ref{dynamical}, the
existence of metastable states has profound consequences in the
far-from-equilibrium, low temperature properties of this system.

\subsection{High temperature properties}
\label{hightemperature}

\begin{figure}
\begin{center}
\includegraphics[width=12cm,height=9cm,angle=0]{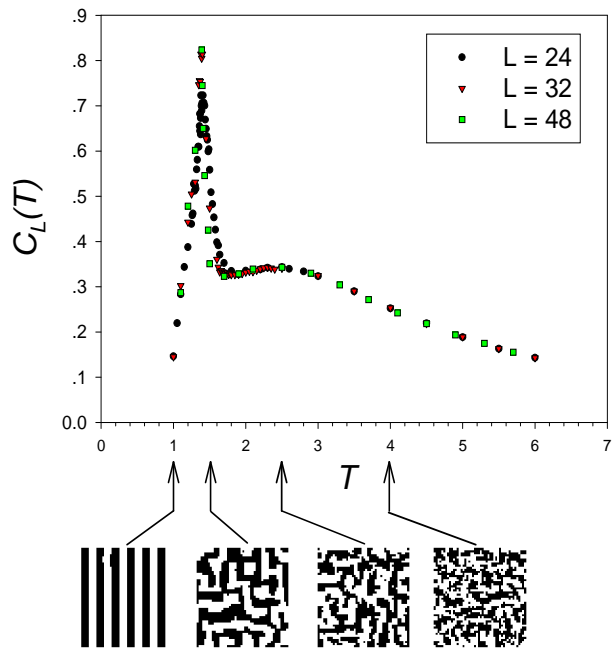}
\caption{\label{calordelta3} Specific heat {\it vs.} $T$  for
$\delta=3$ (corresponding to an h4 ground state) and different
system sizes. Some typical equilibrium configurations at the
indicated temperatures for $L=48$ are shown below. Note the
sequence of transitions  h4 $\rightarrow$ tetragonal $\rightarrow$
paramagnetic.}
\end{center}
\end{figure}
When we increase the temperature the ordered phases undergo a
phase transition into a disordered state. Booth and
coauthors \cite{Booth} showed that above and near the transition
temperature the disordered phase is not paramagnetic, but instead
it consist of extended ferromagnetic domains characterized by
predominantly square corners (see Fig.\ref{calordelta3}). This
phase  presents a fourfold rotational symmetry, as can be observed
in numerical calculations of the structure factor

\[ S(\vec{k}) = \left< \left| \sum_{\vec{r}} \sigma_{\vec{r}} e^{i\vec{k}.\vec{r}}
\right|\right> \]

\noindent which displays four symmetric peaks along the principal
axes of the Brillouin zone \cite{DeBell}.  As temperature is
further increased, the four peaks become gradually smeared into a
circle shaped crown, signaling the continuous replacement of the
fourfold symmetry by the full rotational symmetry of the
paramagnetic phase. Booth and coauthors proposed that the
transition from the striped to this {\it tetragonal} phase can be
characterized as an order-disorder one,  associated to the loss of
orientational order of the striped phase \cite{Booth}. To
characterize this symmetry breaking they introduced an order
parameter

\begin{equation}
m \equiv \frac{n_v-n_h}{n_v+n_h}
\label{orderp}
\end{equation}

\noindent where $n_v$ ($n_h$) is the number of vertical
(horizontal) bonds between spins that are
antiparallel \cite{Booth}.  The absolute value of this order
parameter is one in {\it any} stripe configuration and equals zero
in any configuration with fourfold rotational symmetry.
The numerical simulations performed by Booth and
coauthors of $\left< m \right>$  for values
of $\delta\geq 3$ appeared consistent with a second order
stripe-tetragonal phase transition \cite{Booth}. In the same
parameters region, the presence of the tetragonal phase shows its
signature in the shape of the specific heat curve, as is
illustrated in the example shown in  Fig. \ref{calordelta3}, for
$\delta=3$. We see that the curve presents two peaks. The low
temperature peak increases with the system size $L$ and coincides
with the temperature at which the order parameter (\ref{orderp})
decays.  Thus, it is associated with the stripe-tetragonal phase
transition \cite{Booth}. The temperature of the corresponding
maximum of the specific heat appears to be almost independent of
the system size, a finite size behavior usually observed in
second order phase transitions. The second broader peak at higher
temperature does not depend on the system size and is associated
with the continuous decay of the tetragonal into the paramagnetic
phase. This second peak becomes more pronounced as $\delta$
increases \cite{Booth}.

The existence of the tetragonal phase was first predicted by
Abanov and coauthors \cite{Abanov} in a continuous approximation for
ultrathin magnetic films, and it was only recently verified
experimentally in fcc Fe on Cu(100)
films \cite{Vaterlaus,Portmann}. The work of Abanov {\em et al.}
also predicted that the stripe-tetragonal transition should be either
first order
or the two phases might be separated a third phase characterized
by rotational domain wall defects, that they called an {\it Ising
nematic phase} \cite{Abanov}. Neither the Monte Carlo results of
Booth {\em et al.}, nor the experimental results have shown any evidence
of this phase. However, the results of Booth {\em et al.} appear to be
consistent with a {\it second} rather that a first order
transition, as expected from Abanov {\em et al.} results. Nevertheless,
Booth {\em et al.} pointed out that  the possibility of a weak first
order transition cannot be excluded from their Monte Carlo
calculations. Indeed, extensive Monte Carlo calculations in other
region of the parameters space showed that this is the
case \cite{Cannas1}.
\begin{figure}
\begin{center}
\includegraphics[width=11cm,height=8cm,angle=0]{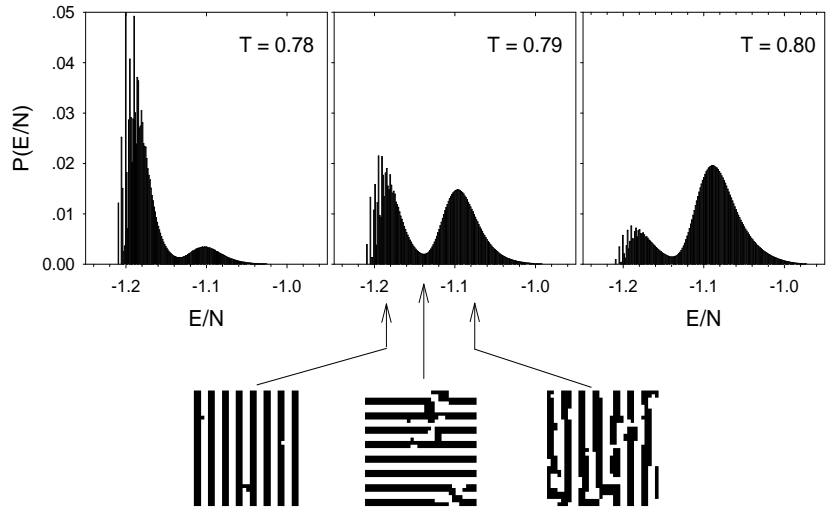}
\caption{\label{hystogram1} Energy per spin histograms for
$\delta=2$, $L=32$ and different temperatures around the
pseudo-critical one $T_c^{(1)}(32) \approx 0.79$. Some typical
equilibrium configurations for the indicated energies are also
shown.}
\end{center}
\end{figure}
Information about the order of the transition can be obtained by
analyzing the energy per spin histograms obtained during a single,
large run of a Monte Carlo simulation for different temperatures,
as in the example shown in Fig.\ref{hystogram1} for $\delta=2$.
The double peak structure of the energy distribution is
characteristic of a first order phase transition. The typical
configurations associated with each peak (i.e., with energies
around the maxima) depicted in Fig.\ref{hystogram1} show that the
low and high energy peaks are associated with the striped ($h=2$)
and the tetragonal phases respectively. The typical configurations
associated with the minimum of the distribution correspond to a
coexistence of both phases. Further evidence of the first order
nature of the transition can be obtained from finite size scaling
properties of the specific heat (\ref{calor1}) and the Binder
fourth order cumulant \cite{Binder}:

\begin{equation}
V_L \equiv 1- \frac{\left< H^4 \right>}{3\left< H^2 \right>^2}
\end{equation}
\begin{figure}
\begin{center}
\includegraphics[width=8cm,height=6cm,angle=0]{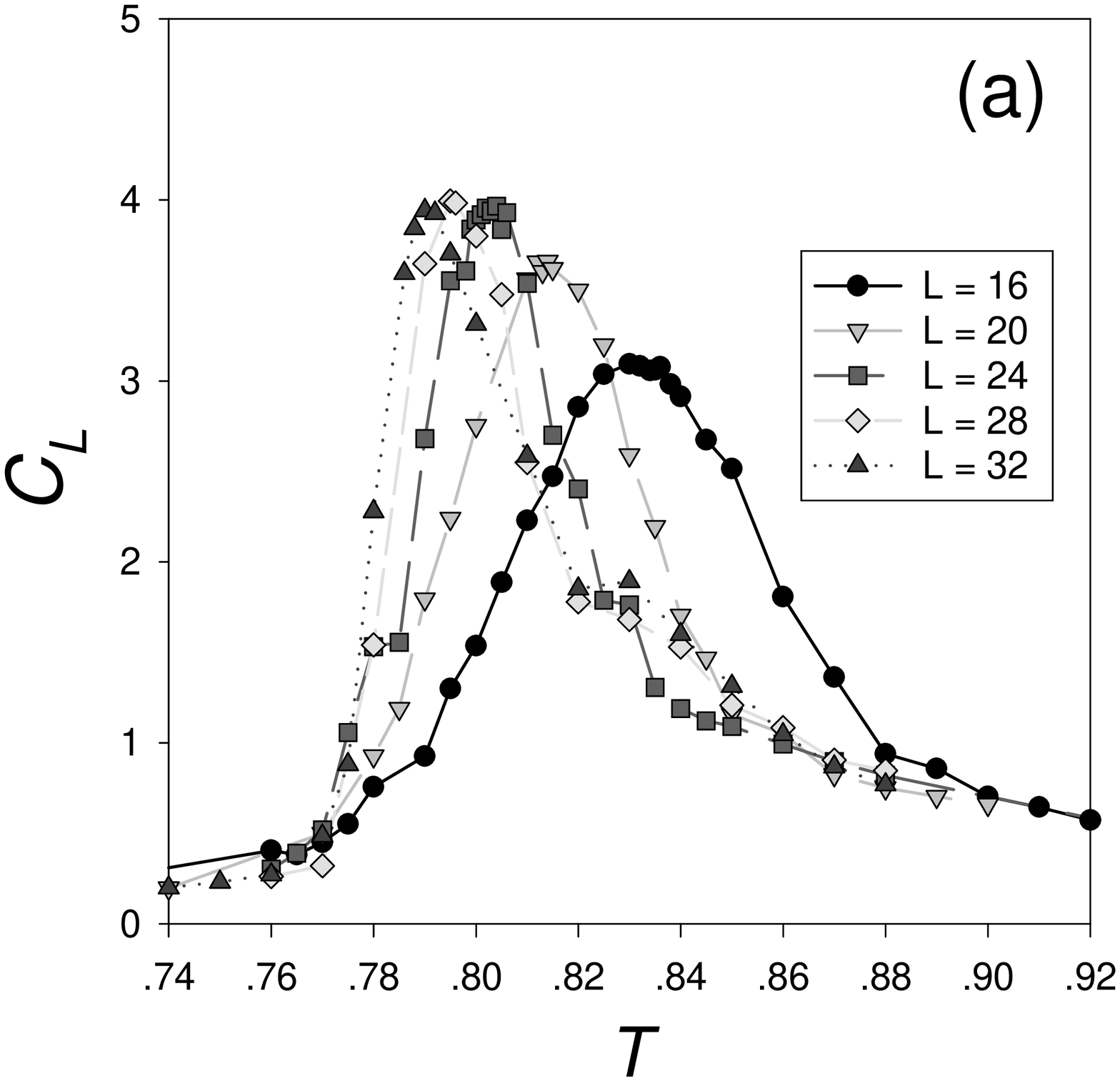}
\includegraphics[width=7cm,height=6cm,angle=0]{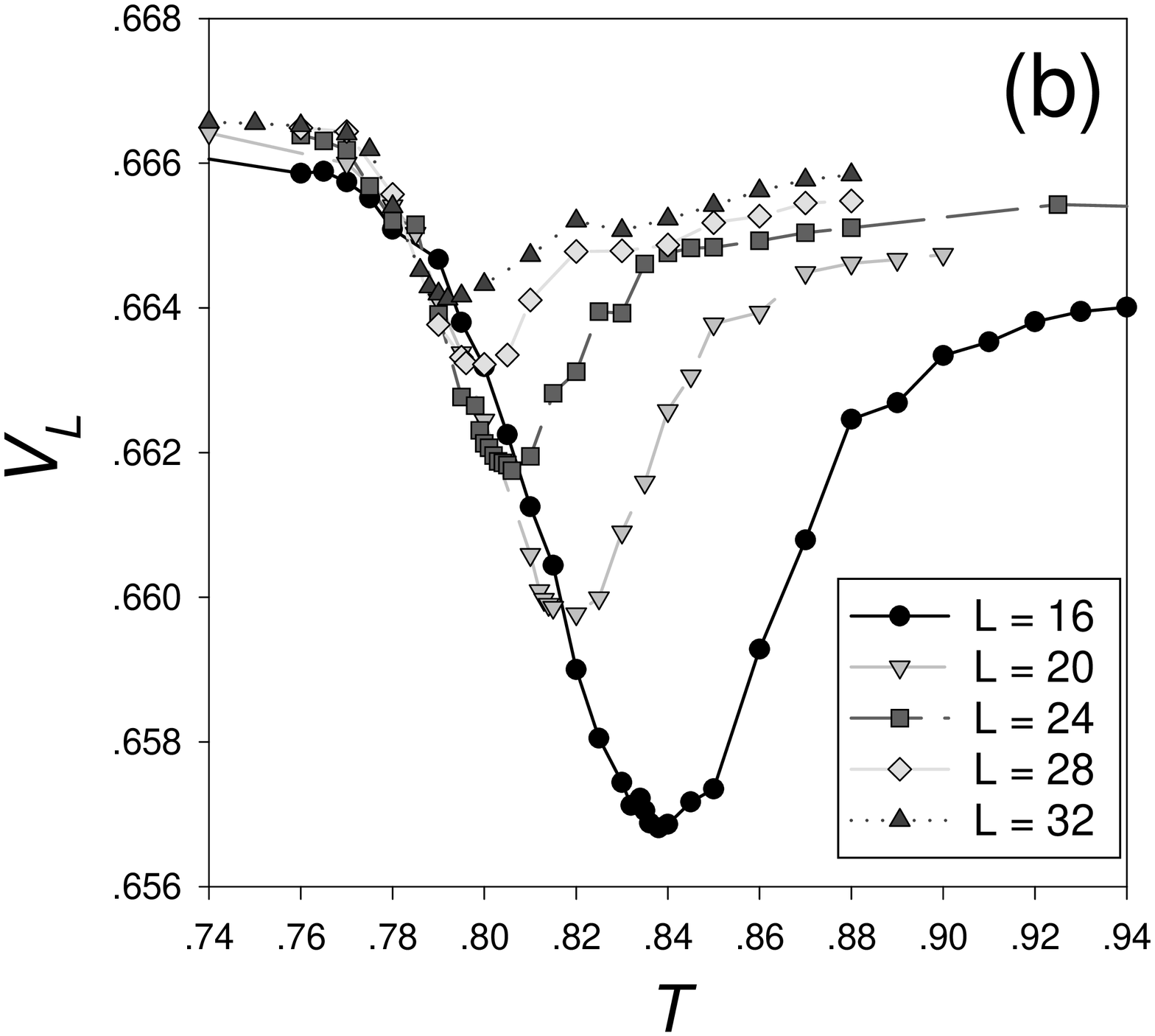}
\caption{\label{momentos} Monte Carlo calculations for $\delta=2$
and different system sizes. (a) Specific heat; (b) Binder
cumulant.}
\end{center}
\end{figure}
\begin{figure}
\begin{center}
\includegraphics[width=8cm,height=6cm,angle=0]{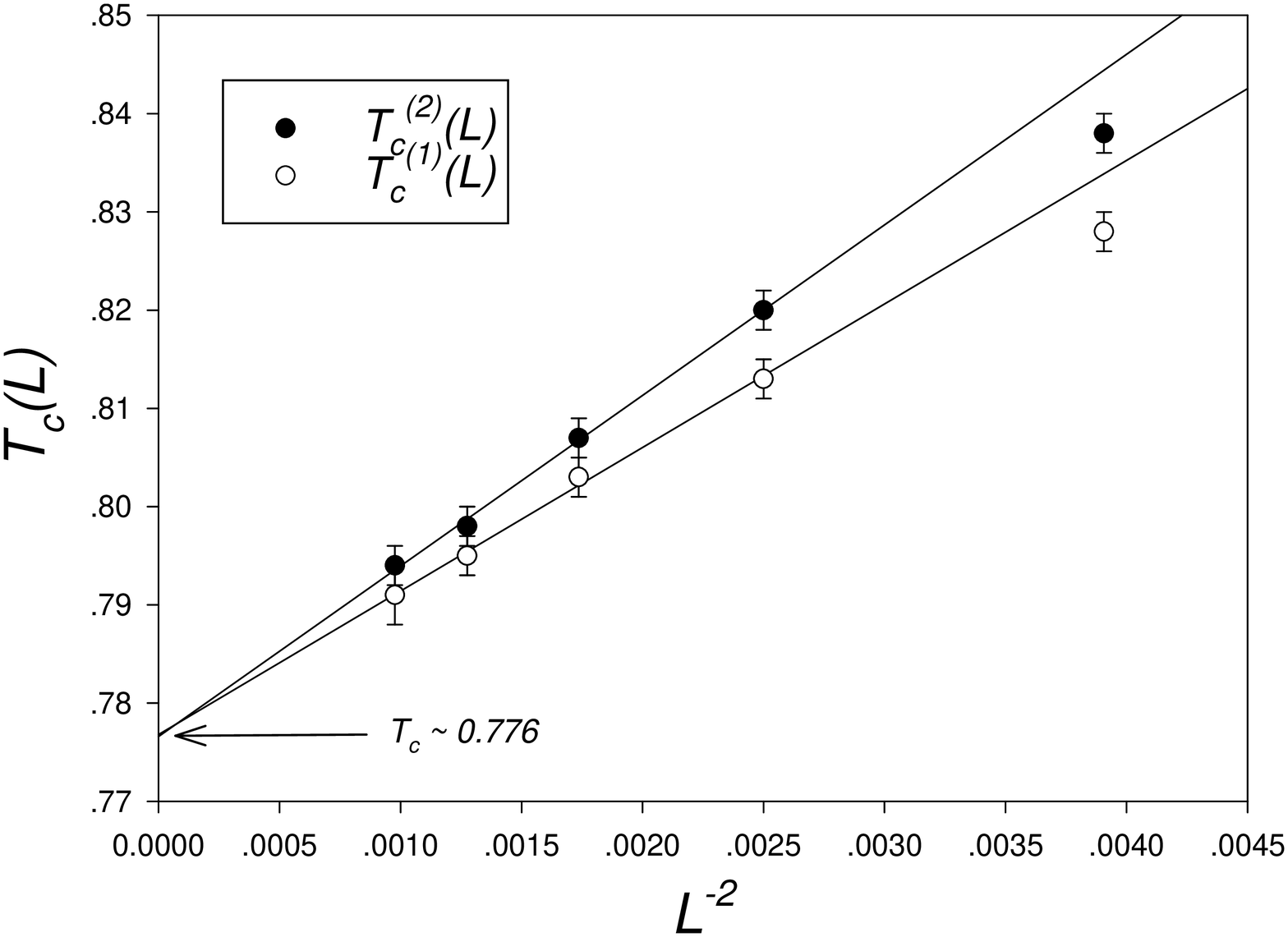}
\caption{\label{pseudo} Pseudo critical temperatures $T_c^{(1)}$
(maximum of the specific heat) and $T_c^{(2)}$ (minimum of the
Binder cumulant) vs. $L^{-2}$ for $\delta=2$.}
\end{center}
\end{figure}
\noindent The last quantity presents a monotonous behavior at the
critical temperature if the transition is continuous \cite{Binder}.
If the transition is first order, $V\rightarrow 2/3$ both for $T
\ll T_c$ and for $T \gg T_c$ (when $L\rightarrow \infty$) and
presents a minimum around
some pseudo-critical temperature $T_c^{(2)}(L)$ \cite{Binder}.
In a first order temperature-driven phase transition the
specific heat presents a
maximum at a different pseudo-critical temperature
$T_c^{(1)}(L)<T_c^{(2)}(L)$. In Fig.\ref{momentos} we see a Monte
Carlo calculation of $C_L(T)$ and $V_L(T)$ for $\delta=2$ and
different system sizes \cite{Cannas1}. The strong dependency of
$T_c^{(1)}(L)$ and $T_c^{(2)}(L)$ on $L$ are characteristic of
first order phase transitions. Moreover, the plot of those
quantities as a function of $L^{-2}$ shown in Fig.\ref{pseudo}
verify the expected finite size scaling behavior in a first-order
temperature driven phase transition \cite{Lee}: $T_c^{(1)}(L)\sim
T_c + A L^{-2}$ and $T_c^{(1)}(L)\sim T_c + B L^{-2}$ with $B>A$,
 $T_c$ being the transition temperature of the infinite system.
\begin{figure}
\begin{center}
\includegraphics[width=12cm,height=9cm,angle=0]{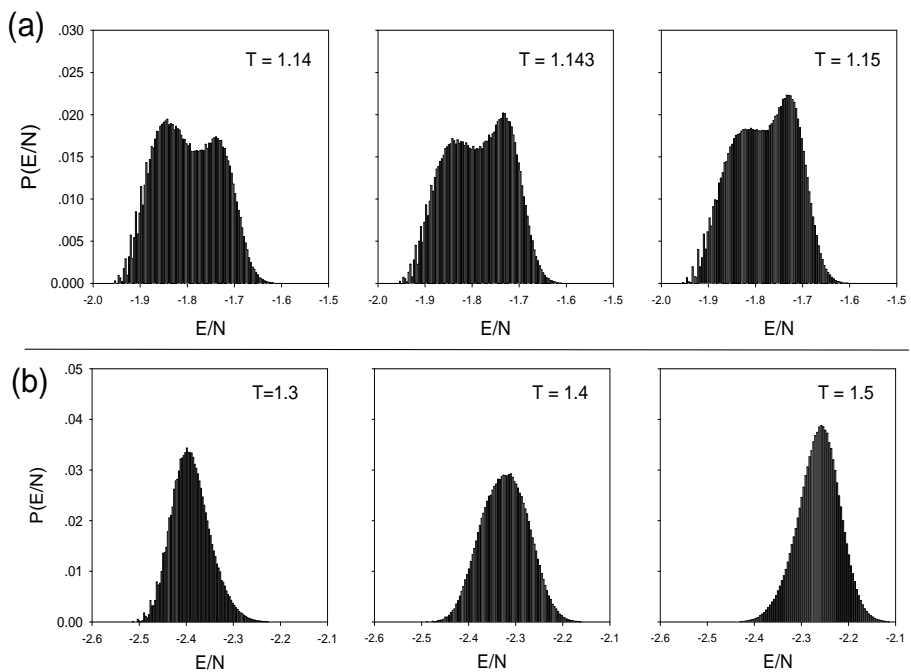}
\caption{\label{hystogram2} Energy per spin histograms for
 $L=24$ and different temperatures around the
pseudo-critical ones $T_c^{(1)}(L)$. (a) $\delta=2.6$; (b)
$\delta=3$.}
\end{center}
\end{figure}

Notice that the internal energies of both phases (roughly
corresponding to the energies of the maxima of the histogram) near
the transition temperature are located very close to each other.
This property is also reflected in the rather shallow shape of
Binder cumulant around $T_c^{(2)}$ (see Fig.\ref{momentos}b) and
evidences the weak nature of the transition. This effects become
more pronounced as $\delta$ increases. Moreover, for values of
$\delta>2.6$ the double peak structure (together with the minimum
of the Binder cumulant) seems to disappear, or at least become
undetectable for small system sizes, as can be seen in
Fig.\ref{hystogram2}. This fact explains the seemingly continuous
nature of the transition observed by Booth {\em et al.}, whose
calculations were performed for  $\delta \geq 3$ \cite{Booth}.
However, analytic calculations on a related continuous model
suggest that it belongs to a large family of systems, or
universality class, in which a first transition for {\it any}
value of $\delta$ is expected on quite general
grounds \cite{Cannas1}. So, the question of the order of the
transition for large values of $\delta$ still remains open.

\section{Dynamical properties}
\label{dynamical}

The dynamics of the two-dimensional Hamiltonian (\ref{Hamilton1})
at low enough temperatures is characterized, as occurs with any
magnetic system that presents long--range order, by the formation
and growth of small regions of ordered spins called {\em domains.}
But unlike other magnetic models without disorder in the
Hamiltonian, in this particular case the competition between the
short range ferromagnetic interactions (that induces the system to
order) and the dipolar antiferromagnetic interactions (that
frustrates it and tends to increases the degeneracy of the ground
state) gives place to an unusual slowing down of the dynamics. And
this peculiar behavior somehow resembles that observed in systems
with imposed disorder (i.e., with randomness in the Hamiltonian),
such as, for instance, Edwards--Anderson spin glasses and random
field Ising models. This is not a minor point, at least from a
statistical mechanics point of view, mainly because there has been
a considerable effort during the last years in trying to find a
lattice model able to catch many of the dynamical and
thermodynamical properties of structural glasses, which also
present a notorious slowing down of the dynamics. And unlike spin
glasses, it is nowadays vastly accepted in the community that an
adequate model for structural glasses can not include disorder in
the Hamiltonian. On the contrary, all the complexity of their
dynamical behavior should be explained only in terms of the
competition between the short range repulsion and long range
attraction character of a some {\em Lennard--Jones}--like
molecular interactions. Although a complete discussion about those
attributes that can make a lattice system an acceptable model for
structural glass is out of the scope of this article, we will
revisit this point below, in order to show some recent evidence
that permit us to suspect that  Hamiltonian (\ref{Hamilton1})
could be considered a good candidate for modelling a bidimensional
structural glass system.

Let us come back to the {\em thin--film} interpretation of
model (\ref{Hamilton1}), which in fact is shared by almost all the
authors who have contributed to its study during the last years.
In 1996, Sampaio, Albuquerque and de Menezes  performed a detailed
study of the relaxation dynamics and hysteresis effects of the
model \cite{Sampaio}. In particular, starting from a fully magnetized
configuration, they numerically analyzed the way the magnetization
relaxes when the magnetic field is suddenly switched off. Inside
the striped phases, where the equilibrium magnetization is
obviously zero, they surprisingly found two different dynamical
regimes, depending on the value of $\delta$. For $\delta \ge
\delta_c \equiv 1.35$ the initially fully magnetized state
quickly relaxes to equilibrium, following the expected exponential
law characterized by a strong dependence on both $\delta$ and
temperature $T$. On the other hand, when $\delta < \delta_c =
1.35$ the relaxation presents a power law behavior with an
exponent that does not depend on $\delta$. This last behavior,
usually found in systems with imposed disorder, motivated new
investigations about the nature of the dynamical behavior of the
model.

Hence, in the next three subsections we will present and discuss those results that have
been found during the last years concerning the out of equilibrium dynamics of
Hamiltonian (\ref{Hamilton1}).

\subsection{Aging}
Aging is one of the most striking features in the off-equilibrium
dynamics of many complex systems. It refers to the presence of strong memory effects
spanning time lengths that in some cases exceed any available
observational time. Unfortunately, the large body of experimental evidence
accumulated on the aging phenomena has not yet lead to a substantial
understanding  of the microscopic principles that can give origin to this
complex phenomenology. Aging phenomena are so relevant mainly because some
kind of {\em universality} appears in its description which makes it possible
to categorize  different systems as members of different classes.
The main feature of aging is the conjunction of extremely slow dynamics
together with non stationary relaxation functions.  In fact, a system that
ages looses the time translation invariance. That means that all
statistical quantities that depend on two times, do not depend only on
their difference, as occurs in an equilibrated state.

Although aging has been seen in a wide variety of contexts and
systems \cite{Bouchaud} (some of them, actually very simple ones
\cite{Ritort}) it is perhaps in the realm of magnetic and
structural glasses dynamics where a systematic study of these
phenomena has been carried out (see  \cite{Bouchaud} and
references therein).

In real materials, aging can be observed, for instance, in a
Thermo--Remanent Magnetization (TRM) relaxation experiment: the
system is abruptly cooled under a magnetic field $\vec{B}$ that is
switched on at $t=0$, when the system is quenched at $T<T_c$, up
to a time $t_w$ when it is suddenly turned off. One then verifies
that the magnetization $M$ at time $\tau +t_w$ depends both on
$\tau$ and $t_w$, with $\lim_{t_w \to \infty} M(t_w+\tau,t_w)
\equiv M(|\tau|)$. One can also measure aging by looking at the
two--time autocorrelation function $C(t,t^{\prime})$. Although it
is hard to measure autocorrelation functions in a real experiment,
it is very simple to calculate it in a numerical experiment. For
an Ising system like the one considered in this article, the
autocorrelation is defined as:
\begin{equation}
\label{correlacion} C(t_w+\tau,t_w) = \frac{1}{N} \sum_{i=1}^N
S_i(t_w+\tau) S_i(t_w)
\end{equation}
In these numerical experiences it is not necessary  to consider an
external magnetic field. The system is simply quenched at time
$t=0$ from a very high temperature (usually infinite temperature)
below the transition temperature and let to evolve during certain
{\em waiting time} $t_w$, when the configuration $\{S_i(t_w)\}$ is
stored. From then on the two time autocorrelation function
(\ref{correlacion}) is calculated. If the system attains an
equilibrium state, then $C(t_w+\tau,t_w)= C(\tau)$. But, if the
system does not equilibrate in a reasonable time, then $C$ will
explicitly depend both on $\tau$ and $t_w$, indicating the
presence of aging.

There are basically two scenarios within which aging can
emerge. On one hand, aging appears as a
consequence of weak ergodicity breaking \cite{Bouchaud} and it is
related to the complex structure of the region of phase
space that the system explores in time. This is the case, for
instance, in the Sherrington-Kirkpatrick (SK) \cite{Sherrington} model and other
spin glass models in which the complexity of the energy function
is associated to a certain degree of randomness and/or frustration
in the Hamiltonian.
On the other hand, the onset of aging in many systems
derives from the presence of coarsening processes that give place
to a drastic slowing down of the dynamics. In this case the scaling
law of the two-time autocorrelation function is ruled by the
following expression
\begin{equation}
C(t+t_w,t_w) \sim f \bigl( L(t)/L(t_w) \bigr),
\end{equation}
where $L(t)$ is the mean linear size of the domains at time $t$
\cite{Bray}.

\begin{figure}
\includegraphics[width=9cm,height=6cm,angle=0]{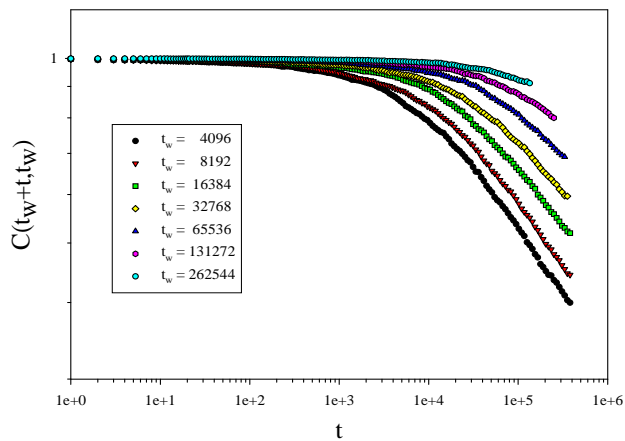}
\caption{Autocorrelation function $C(t_w+t,t_w)$ vs. $t$ for
$\delta = 2.0$,  $T=0.25$ and different waiting times $t_w$.}
\label{aging-fig1}
\end{figure}

In Fig. \ref{aging-fig1} we see a typical behavior of
$C(t_w+t,t_w)$ inside the striped phase for $\delta=2.0$, $T=0.25$
and different values of $t_w$. Clearly we verify the presence of
aging, identified by the strong dependence on both $t$ and $t_w$.
For a given $t_w$ we observe that the system stays in a
quasi--equilibrium state before $C(t_w+\tau,t_w)$ finally relaxes
to zero. Furthermore, as the time the system spends in the ordered
phase $t_w$ increases, the time the system spends in the
quasi--equilibrium state also increases (this phenomenon is the
origin of the name {\em aging}).

An insight about the nature of the aging phenomena observed inside
the striped phase can be obtained by analyzing the scaling
properties of $C(t+t_w,t_w)$ in different regions of the phase
diagram, which can tell us about the universality class of the
dynamical regime. Indeed, there seems to be two different
dynamical regimes \cite{Toloza}, depending on the value of
$\delta$. When $\delta>\delta_c$ and $T=0.25$, $C(t+t_w,t_w)$
displays the scaling form:
\begin{equation}
C(t+t_w,t_w) \propto f\left(\frac{t}{t_w}
\right).
\end{equation}

\noindent while for $\delta<\delta_c$ at the same temperature  the
results seems to be consistent with a logarithmic scaling law of
the form~\cite{Toloza}:
\begin{equation}
C(t_w+t) \propto g\left(\frac{\ln{(t)}}{\ln{(t_w)}} \right).
\end{equation}

Let us extract now some conclusion of these results. If the
slowing down of the dynamics is ruled by the domain growth
process, as expected in this kind of model without disorder,
after a certain time $t$ one can measure the average linear domain
size $L(t)$. The time evolution of quantities like the
autocorrelation function will then present a crossover from
the dynamical regime characterized by length scales smaller
than the domain size $L(t_w)$ to a regime at larger
scales, dominated by domain growth through the movement of
domain walls. In this scenario, scaling arguments lead to the
following (and already mentioned) dependency of $C(t_w,t_w+t)$
\begin{equation}
C(t_w,t_w+t) \propto \Theta  \left(\frac{L(t+t_w)}{ L(t_w)}\right)
\end{equation}
Hence, when $t \gg t_w$
\begin{equation}
C(t_w,t_w+t) \to \Theta\left(\frac{L(t)}{L(t_w)} \right)
\end{equation}
is expected.

Summarizing, the results for $\delta > \delta_c$ are consistent
with an algebraic growth of the linear domain size of the form
$L(t) \propto t^{\psi}$. But what emerges as a much curious result
is the behavior observed for $\delta < \delta_c$, which appears to
be consistent with a logarithmic time dependence of the average
domain size $L(t) \propto (\ln{(t)})^{\psi}$ predicted by an
activated dynamic scenario proposed by  Fisher and Huse
\cite{Fisher} in the context of spin glasses, in which both
disorder and frustration generate active droplets excitations with
a broad energy distribution.

Another consequence of the loose of the time invariant translation
of a system that ages during its relaxation, is the violation of
the Fluctuation Disipation Theorem (FDT). Let us suppose that an
inhomogeneous external magnetic field $h_i(t)$ is switched on at
time $t_w$; the conjugate response function at time $t_w+t$ can be
defined as
\begin{equation}
R(t_w+t,t_w) = \frac{1}{N} \sum_{i} \frac{\partial \left<
S_i(t_w+t)\right>}{\partial h_i(t_w)}
\end{equation}

\noindent Let us now assume that the system has attained true
thermodynamical equilibrium. Then, time translational invariance
holds and both the autocorrelation and the response functions
depend only on the time difference $t$ and  the FDT gives us a
precise relationship between $R(t)$ and $C(t)$:

\begin{equation}
R(t) = \frac{1}{T}  \frac{\partial C(t)}{\partial t}.
\end{equation}

For a system that ages, as the one studied in this paper, one can
identify two different regimes, as we have already seen: for small
values of $t$ ($t/t_w \ll 1$) the system is a quasi--equilibrium
states, at least for large $t_w$, and the all the equilibrium
properties hold. But in the aging regime, when $t\gg t_w$, both
the time translation invariance and the Fluctuation Dissipation
Theorem do not hold. In other words, $C(t_w+t,t_w)$ and
$R(t_w+t,t_w)$ depend explicitly on $t$ and $t_w$. In particular,
for a great variety of disordered models and also during some
coarsening processes, in this last regime a generalized version of
the $FDT$ holds, which asserts that
\begin{equation}
R(t_w+t,t_w) = \frac{X(t+t_w,t_w)}{T}  \frac{\partial C(t+t_w,t_w)}{\partial t_w}
\end{equation}
with $X(t+t_w,t_w) \neq 1$. Moreover, for large enough values of
$t_w$, $X(t+t_w,t_w)$ becomes a function of time only though
$C(t_w+t,t_w)=X(C(t+t_w,t_w))$. The function $T/X(C)$ can be now
interpreted as a non--equilibrium {\em generalized temperature}.
This kind of analysis has also been applied in \cite{Stariolo}
to model (\ref{Hamilton1}). Surprisingly, it was found that the
system behaves in the same way, irrespectively of the value of
$\delta$. Actually it was observed that, as $t_w \to \infty$, the
generalized temperature tends to zero, which is a clear signature of
a coarsening process.

\subsection{Coarsening}
As we have already seen, when a magnetic system is suddenly
quenched from a very high temperature (above the critical
temperature $T_c$) into the ordered phase, small clusters of
ordered spins form which are usually called {\em domains}.
Immediately after the cooling the system finds itself in a
disordered state induced by the abrupt quench, but accordingly
to the thermodynamical laws, it would  evolve into an ordered
state. Under these circumstances the small domains of ordered
states start growing, in a process that is  usually called {\em
coarsening}. If we could take a snapshot of  the system a few
seconds after the quench, we would identify a  patchwork of
small domains, ordered in so many different forms as different
ground states the system admits. But, since there is a
interfacial free energy cost associated with the surface of the
boundary between domains, as the domains grow they compete with each
other in order to impose their own order. As a consequence of this
competition, the patchwork of domains evolves in time, trying to
decrease the free--energy of the whole system by decreasing the
area of the interface region between domains.

Coarsening is an ubiquitous phenomenon in nature  that largely
exceeds the realm of magnetism \cite{Sethna}. Even more, coarsening plays a fundamental
role
in many  industrial processes, as for instance in molten iron.
When the molten iron is suddenly quenched below the melting
temperature, the originally dissolved fraction of carbon present
in the system precipitates out. If the quench is fast enough,
the carbon can not float to the top but stays dispersed through
the iron forming small particles, and many of the relevant properties of
cast iron depends specifically on the size and form of the carbon
small clusters. Hence, controlling the quality of the iron requires
a precise knowledge of the coarsening process of carbon particles.
Other examples of coarsening in physical systems appear in foams,
the ordering of binary alloy following a quench from above to
below its order--disorder transition temperature and the phase
separation of a binary fluid following the quenching from the
one--phase to the two--phase region of its phase diagram.

The simplest way to characterize the coarsening process is by
measuring the time evolution of the {\em characteristic domain
linear size} $L(t)$. This procedure, which is hard to carry
out in an experiment, can be easily implemented in numerical
simulations of Ising models like the ones presented in this
article. Coarsening has been vastly studied during the last
decades, both experimentally and theoretically, but there are
still many open questions, especially concerning  the role of
frustration in the time evolution of $L(t)$.  The domain growth
scenario in the ferromagnetic phase of the Ising model is very
much alike the critical dynamics observed in the neighborhood of a
critical point. The competition among the different domains
trying to impose their order to the whole system (as discussed
above) yields to a power law behavior of the average linear size
$L(t)$
\begin{equation}
\label{eq1} L(t) \propto t^n
\end{equation}
Note that this critical--like slowing down of the dynamics occurs even
far away from the  critical point. Since the domain linear size $L(t)$ will
eventually reach macroscopic length scales one can expect that
the coarsening exponent $n$ will be independent of many of the
microscopic details.  In other words, there should be only a few
dynamical {\em universality classes}, as in fact occurs in
the study of critical phenomena.
Two of these universality classes are already well understood. The
first class corresponds to the case of a microscopic dynamics that
preserve the order parameter of the system, and they are
characterized by an exponent $n=1/3$. They are usually called {\em
Lifshitz--Slyozov} \cite{Lifshitz1} growth processes, and a typical
realization is the spinodal decomposition. The second class is
associated to those microscopic dynamics for which the order
parameter is not conserved, and they are characterized by an
exponent $n=1/2$. As we indicated in
\ref{simulaciones}, our simulations were all performed using a
heat bath or Metropolis Monte Carlo dynamics which correspond to
this second universality class, usually called {\em
Lifshitz--Allen--Cahn} or {\em Curvature--driven}
growth \cite{Lifshitz2,Allen2}. A typical physical example is given
by the process of grain growing in metals.

But, there is also a third universality class observed in systems
with imposed disorder, i.e., in systems whose Hamiltonian
includes, besides the dynamical variables, a set of random
variables necessary to describe the structural randomness of the
system. In these cases the coarsening process departs from the
usual power law of Eq. (\ref{eq1}) and is characterized by a
logarithmic growth rule of the form:
\begin{equation}
\label{eq2}
L(t) \sim  \ln{(t)}
\end{equation}
Spin glasses, random field models and models with random quenched
impurities, are all examples of this third universality class.
During many years it was not clear whether this behavior was
restricted to systems with imposed disorder. Nowadays it is believed that the
departure from the power law dynamics is related to the existence
of free energy barriers that the coarsening system has to overcome
which grow linearly with $L(t)$ \cite{Shore}. In fact, there are
also a few examples of systems with logarithmic coarsening that do
not present randomness in the Hamiltonian, as for instance the
three dimensional Ising model with nearest--neighbors
ferromagnetic interactions and next--nearest--neighbors
antiferromagnetic interactions, whose dynamics has been described
by Shore, Holzter and Sethna in \cite{Shore} (from now on we will
refer to it as the Shore model, even when it has already been
analyzed by other authors\cite{Oitmaa}). What seems to be the fundamental
ingredient for the appearance of logarithmic coarsening is
existence of a certain degree of frustration in the microscopic
interactions.

When coarsening is analyzed at zero temperature, it is relatively
easy to determine numerically the domain growth law. But when the
system is analyzed at a non zero temperature new challenges
emerge. In particular, it is very hard to distinguish real domain
structures from occasional small clusters generated by thermal
fluctuations. One way to overcome this difficulty is to use a
technique based on the spreading of damage method proposed by
Derrida \cite{Derrida} and later on improved by Hinrichsen and
Antoni \cite{Hinrichsen}.  The method compares, as time goes on,
the state of a suddenly quenched system with replicas of the same
system initialized in the corresponding ground states. In
particular, one needs to take into account the time evolution of
so many replicas as possible ground states the system admits.
Remember that in our case, the system has, depending on the value
of $\delta$, $4h$ possible ground states, which makes it very
difficult to apply this technique for large values of $\delta$.
Both the quenched system and all its replicas evolve under
the same thermal noise or, in other words, using the same random
sequence in the updating process. All those spin flips that occur
simultaneously in the quenched replica and in one of the ground
state systems are considered thermal fluctuations, and they are
not taken into account in the calculus of the domain areas.

We will summarize now the results of the coarsening dynamics of
Hamiltonian (\ref{Hamilton1}), which are presented in detail in
\cite{Gleiser2}. We focus our analysis in the particular case in which
the system is quenched into the $h1$ phase for values of $0.8 \le
\delta \le \delta_{1,2}$ (see Fig. \ref{phased}). To characterize
 the domain growth, the domain areas
$A(t)$, defined as the number of spins inside each ordered cluster
are first calculated. Then, the characteristic linear domain size
is calculated as $L(t) = \sqrt{A(t)}$. Note that in this
particular case, the spreading of damage method required a
simultaneous comparison among five different replicas of the
system (one for the quenched replica plus four for each possible
ground state).

\begin{figure}
\includegraphics[width=9cm,height=6cm,angle=0]{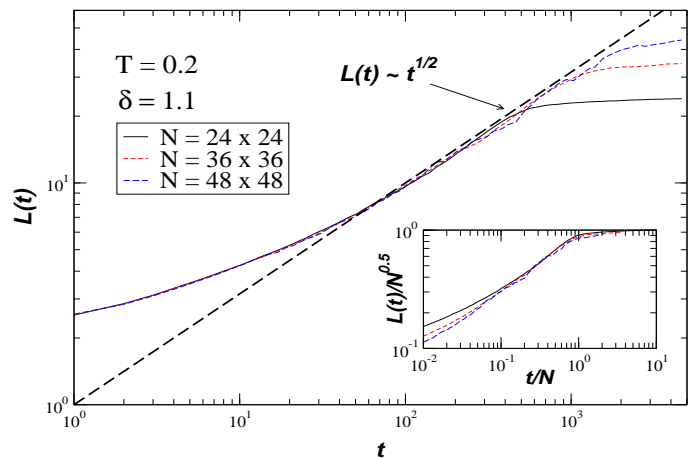}
\caption{Characteristic linear domain size $L(t)$ as a function of
$t$ for for $\delta=1.1$, $T=0.2$ and three different system
sizes. The dashed line indicates the curve $L(t) \sim t^{1/2}$.}
\label{coarsen-fig1} \vspace{0.7cm}
\end{figure}

In Fig. \ref{coarsen-fig1} we present the behavior of the
characteristic linear domain size $L(t)$ when $\delta=1.2$ and
$T=0.2$, for three different system sizes $N=L\times L$ with
$L=24$, $36$ and $48$. Remember that in this
point of the phase space there are no metastable states (see Fig.
\ref{phased}). We observe that after a very short transient the
system enters into the expected coarsening regime $L(t) \sim
t^{1/2}$, as corresponds for a dynamics that does not preserve the
order parameter. For large times,  we see that $L(t)$
saturates in a value that coincides with $N^{1/2}$, indicating
that the domain sizes have reached the size of the system. However, in
\ref{hightemperature} we have seen that inside the ordered phase
one can identify two different situations. In particular, for
$\delta=1$ we known that at low enough temperatures (actually for
$T < 0.4$) there are also metastable states corresponding to
stripes of width $h=2$.

\begin{figure}
\includegraphics[width=9cm,height=6cm,angle=0]{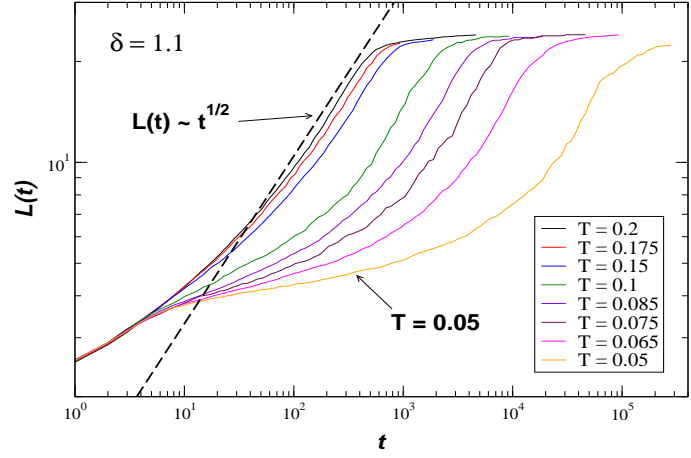}
\caption{Characteristic linear domain size $L(t)$ as a function of $t$ for
for $\delta=1.1$ and $N=24 \times 24$ and eight different temperatures
indicated in the figure.}
\label{coarsen-fig2}
\end{figure}

In order to determine the effect of these metastable states in the
domain dynamics of the model, let us consider what happens when
 the final temperature of the quenching is lowered. In Fig.
\ref{coarsen-fig2} we show the time evolution of $L(t)$ when
$\delta=1$ and $N=24 \times 24$ for eight different temperatures,
ranging from $T=0.2$ to $T=0.05$. What emerges from this analysis is
the appearance of a new dynamical regime associated with the
existence of metastability. When $0.1 < T < 0.2$ the system
displays always the same behavior described in Fig.
\ref{coarsen-fig1}. But, for $T \le 0.1$ a regime of slow growth
develops at intermediate time scales before the system crosses
over to the $t^{1/2}$ coarsening regime. As we  lower the
temperature further, this intermediate regime extends to larger time scales;
nevertheless, all the curves eventually cross over to the $n=1/2$ regime.
Note that for short times there seems to be a change in the
concavity of the curves $L(t)$ when the final quenching
temperature crosses the value $T=0.1$, which coincides, for
$\delta=1.1$ with the spinodal boundary of the metastable phase of
width $h=2$ inside the equilibrium phase of width $h=1$.

\begin{figure}
\hspace{1cm}\includegraphics[width=8cm,angle=270]{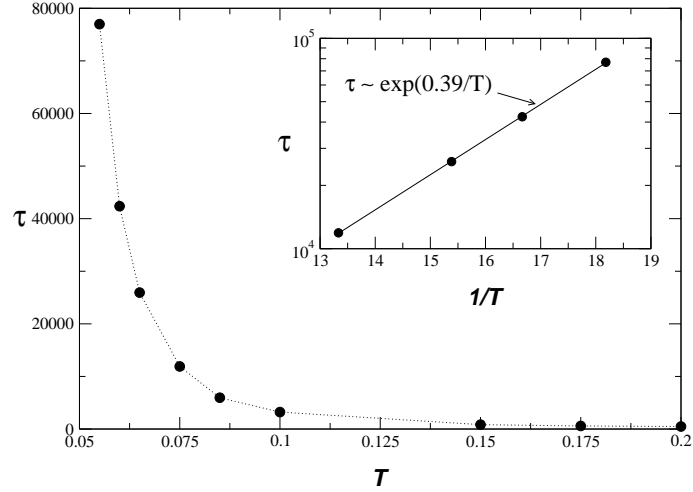}
\caption{\label{coarsen-fig3} Crossover time $\tau$ vs. $T$ for
$\delta=1.1$ and $N=24 \times 24$. On the inset, an Arrhenius plot
of $\tau$ vs. $1/t$ for the four lower temperatures showed in Fig.
\ref{coarsen-fig3}. The straight line in the inset indicates the
best fitting.}
\end{figure}

 To  characterize further the behavior of the linear domain
size $L(t)$ let us consider  the crossover time $\tau$, given by
the intersection of the power law branch of the curve and the
horizontal saturation branch. In Fig. \ref{coarsen-fig3} we show a
plot of $\tau$ as a function of $T$. While for temperatures greater
than $T=0.1$ the crossover time presents a  linear dependency
with $1/T$, in the region of metastability $T \le 0.1$ we observe
an exponential increasing of $\tau$ as $1/T$ increases, as can be
deduced from the Arrhenius plot presented in the inset of Fig.
\ref{coarsen-fig3}. The straight line denotes the best linear
fit, given by
\begin{equation}
\label{eq3} \tau = 62.5 \, \exp{(0.39/T)}
\end{equation}

\begin{figure}
\includegraphics[width=12cm,angle=0]{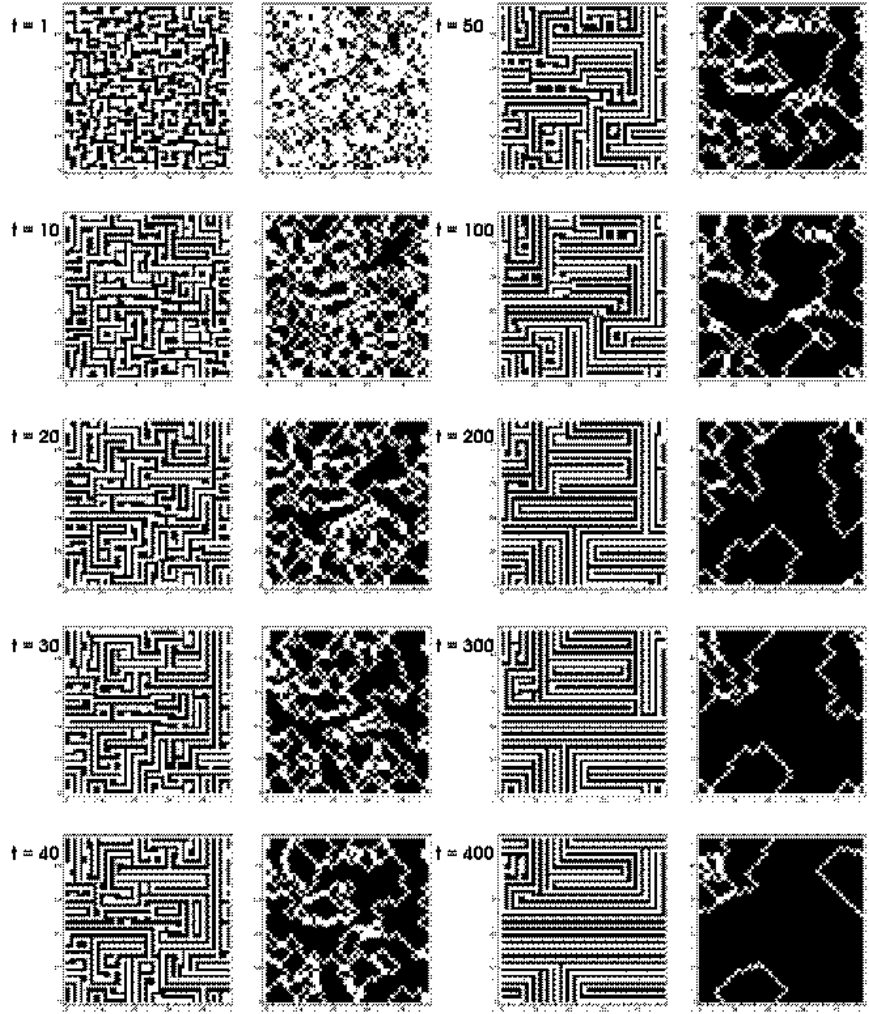}
\caption{Snapshots of the coarsening process when a system with
$\delta=1.1$ and $N=48 \times 48$ spins was quenched to T=0.2}
\label{coarsen-fig4}
\end{figure}

In Fig. \ref{coarsen-fig4} we present a temporal sequence of snapshots of a
system of $L=48$ spins and $\delta=1.1$ at $T=0.2$, where the system
does not present the slow intermediate regime.  The pictures appear in pairs.
The left pictures correspond to configurations of the system at different times,
where black points indicate spins up and white points spins down. In the
right pictures we can visualize the corresponding domain walls obtained with the method of
Hinrichen and Antoni \cite{Hinrichsen}, previously described. Here the black points
represent
spins inside a domain while the white points indicate spins that belong to
the domain wall. Starting from the left top the snapshots correspond to
time steps $t=1$, $10$, $20$, $30$, $40$, $50$, $100$, $200$, $300$ and
$400$. After a few time steps, certain degree of local order is achieved, and
we can identify small clusters of striped structures. As the system evolves
small domains appears, which can be identified in the right pictures as
black areas (see $t=30$ and $t=40$). These small domains start to grow
and the coarsening process can be easily observed (compare $t=50$ and $t=100$).
Note that some domain walls are still very thick, and seem to persist longer
times. The domain walls tend to align in a diagonal direction. We can see that once a
thick domain wall disappears the domain boundary around
it quickly moves. In other words, thick walls seem to pin the motion of
thin walls.

\begin{figure}
\includegraphics[width=12cm,angle=0]{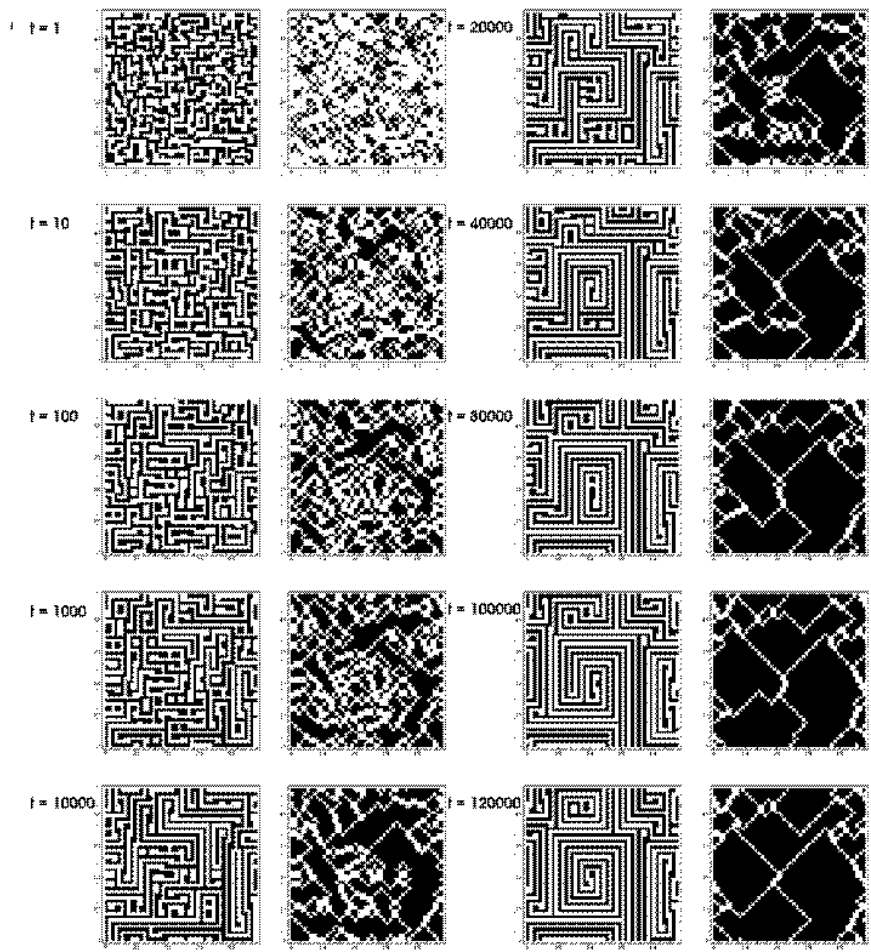}
\caption{Snapshots of the coarsening process when a system with
$\delta=1.1$ and $N=48 \times 48$ spins was quenched to T=0.05}
\label{coarsen-fig5}
\end{figure}

Finally, in Fig. \ref{coarsen-fig5} we present a new sequence, but now
obtained by suddenly quenching a system with $\delta=1.1$ and $N=48\times 48$
into the metastability region, at $T=0.05$. Note that the first
five snapshots are similar to the first five snapshots of the coarsening
process at $T=0.2$. Small domains appear of stripes of width $h=2$
(compare the time scales of this figure and the former). In the
high temperature coarsening this small blocks do not seem to play
a fundamental role in the growth of the domains. However, in the
snapshots of the low temperature coarsening it is clear that this
blocks slow the domain growth. Even more, these blocks seem to
pin the domain walls and slow the dynamics. In other words, the
crossover from the slow logarithmic regime to the $t^{1/2}$
regime will be characterized by the time needed to depin this
blocks and free the domain walls.
These dynamical behaviors present a strong resemblance with the ones observed
 in the two dimensional Shore model \cite{Shore}.
The presence of NNN antiferromagnetic bonds in this model introduce free--energy
barriers to domain coarsening that are independent of the domain size
$L$\cite{Shore}.
Such barriers in this model are a consequence of a corner rounding
process which generates structures that block the coarsening
dynamics\cite{Shore}.
Hence, the system is stuck and coarsens little on time
scales $t \ll \tau_B(T) = exp{(F_B /T)}$ ($F_B$ being the height of
 the barrier), while on time scales $t \gg \tau_B(T)$ the free--energy barrier
 can be crossed and the $t^{1/2}$ behavior emerges. In fact,   the time to completely
shrink  squares of h2 phase immersed in an h1 phase coincides with the divergence observed
in the crossover time from the slow growth to the $t^{1/2}$ regime \cite{Gleiser2}. These
results are consistent with the presence of
free--energy barriers independent of the domain size $L$,
associated with blocking clusters of the metastable phase,
which generates a crossover in the coarsening behavior as
we cross the spinodal line. Since the cross over time diverges
as the temperature is lowered, the very slow behavior at intermediate
times may be indistinguishable from a logarithmic law.

\subsection{Slow cooling}

As we have already seen, at very low temperatures where
metastability effects appear, the system displays an intermediate
slow regime. Nevertheless, after certain period of time that
diverges as $T \to 0$, it finally enters into the expected
algebraic regime $L(t)\propto t^{1/2}$. In this subsection we
present another interesting way of analyzing the out of
equilibrium dynamics of the model. Starting from a thermalized
state above the transition temperature, we measure the average
domain linear size $L(t)$ when the temperature is slowly lowered
below the transition temperature.

To simulate the slow cooling, the  procedure is as follows \cite{Gleiser2}: the
system is initially thermalized at certain temperature $T_0$
above the transition between the ordered and the disordered phases
and then we lower the temperature during the Monte Carlo
simulation at a constant rate
\begin{equation}
T = T_0 - rt \, ,
\end{equation}
where $r$ denotes the constant cooling rate of the experiment.
Once the system has reached the state with $T=0$  one calculates the
average linear domain size $L_0$ at the end of the cooling
process. Since the dynamics gets trapped by the coarsening
process, the system will never reach the ground state for any
non--zero values of $r$. Furthermore, $L_0$ strongly depends on
the cooling rate $r$ and on the coarsening universality class of
the system \cite{Lipowski}. For a glassy system, characterized by the logarithmic
domain growth law (\ref{eq2}) the relationship between $L_0$ and
$r$ is also logarithmic:
\begin{equation}
L_0 \sim -\ln{(r)}
\end{equation}
while for a system  that follows  an algebraic coarsening law one
finds a much faster growth of the form:
\begin{equation}
L_0 \sim r^{n},\ ,
\end{equation}
with $n=1/2$ or $n=1/3$ depending on the particular dynamics.

\begin{figure}
\includegraphics[width=9cm,height=6cm,angle=0]{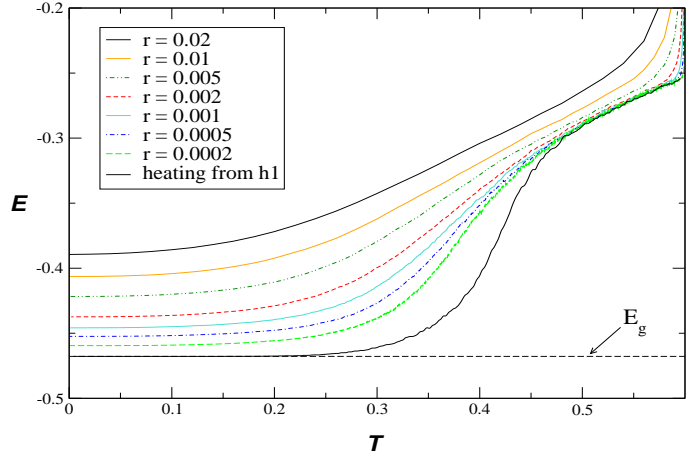}
\caption{Internal energy $E$ as a function of temperature $T$ for
seven different cooling rates, (from top to bottom) $r=0.02$,
$0.01$, $0.005$, $0.002$, $0.001$, $0.0005$ and $0.0002$, for
$N=64 \times 64$ and $\delta = 1$. The lowest curve corresponds to
a heating process from the ordered states (striped state with
$h=1$). The dashed line indicates the ground state energy
$E_g=-0.467758$.}
 \label{cooling-fig1}
\end{figure}

Along the simulations one can also monitor the energy excess $\delta E(r)
= E_0(r) - E_g$ where $E_0(r)$ is the final value of the energy when
the system is cooled at rate $r$ and $E_g$ is the energy of the ground
state. In Fig. \ref{cooling-fig1} we show the temperature dependence
of the internal energy $E$ for seven different cooling rates
$r=0.02$, $0.01$, $0.005$, $0.002$, $0.001$, $0.0005$ and $0.0002$, for
a system of size $N= 64 \times 64$ and for $\delta = 1.0$ (inside the
striped phase of width $h=1$).

\begin{figure}
 \vspace{1.2cm}
\includegraphics[width=9cm,height=6cm,angle=0]{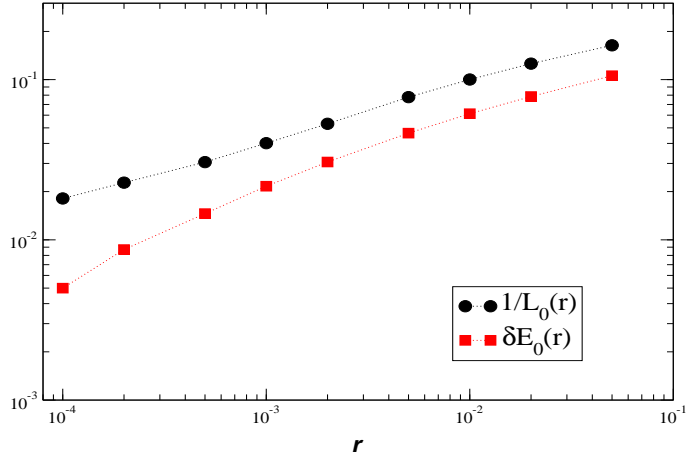}
\caption{Excess energy $\delta E$ and $1/L_0(r)$ as a function of
the cooling rate $r$ for $\delta=1.0$.} \vspace{0.3cm}
 \label{cooling-fig2}
\end{figure}

In Fig. \ref{cooling-fig2} we plot both $1/L_0(r)$ and $\delta E_0$ as a
function of the cooling rate $r$. The quantity $L_0(r)$ was calculated
by using the method proposed by Cirilo {\em et al.} in  \cite{Cirilo}.
The best fits gave for both quantities a power law behavior
\begin{equation}
\delta E(r) \propto r^{0.35}  \qquad \text{and} \qquad L_0(r)=1.9^{-0.37}
\end{equation}
which can not be interpreted neither as a logarithmic nor as the
$n=1/2$ algebraic law. But this behavior can be easily understood:
when the system is slowly cooled from the disordered  phase into
the metastability region, it must pass through the fast dynamics
temperature range, which allows the system to form certain local
order at a fast rate, before it enters into the slow intermediate
regime.

\begin{figure}
\includegraphics[width=9cm,height=6cm,angle=0]{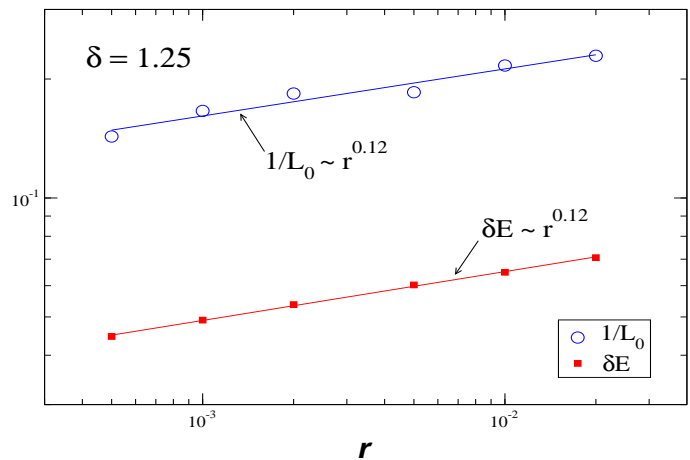}
\caption{Excess energy $\delta E$ and $1/L_0(r)$ as a function
of the cooling rate $r$ for $\delta=1.25$. }
\label{cooling-fig3}
\end{figure}

To verify the former scheme let us consider the slow cooling of a
system with $\delta=1.25$, which is near the border between the
$h=1$ and $h=2$ striped phases. We chose this value because in
this particular case the system passes directly from the
disordered phase into the metastability region without entering in
a fast dynamics regime (see Fig. \ref{phased}). In Fig.
\ref{cooling-fig3} we plot again both $1/L_0$ and $\delta E$ vs.
$r$ for $\delta =1.25$. The behavior of $L_0$ can be well fitted
by a power law with a small exponent
\begin{equation}
L_0(r) \propto r^{-0.12}.
\end{equation}
which is very hard to distinguish from a logarithmic law.

\begin{figure}
\includegraphics[width=6cm,angle=0]{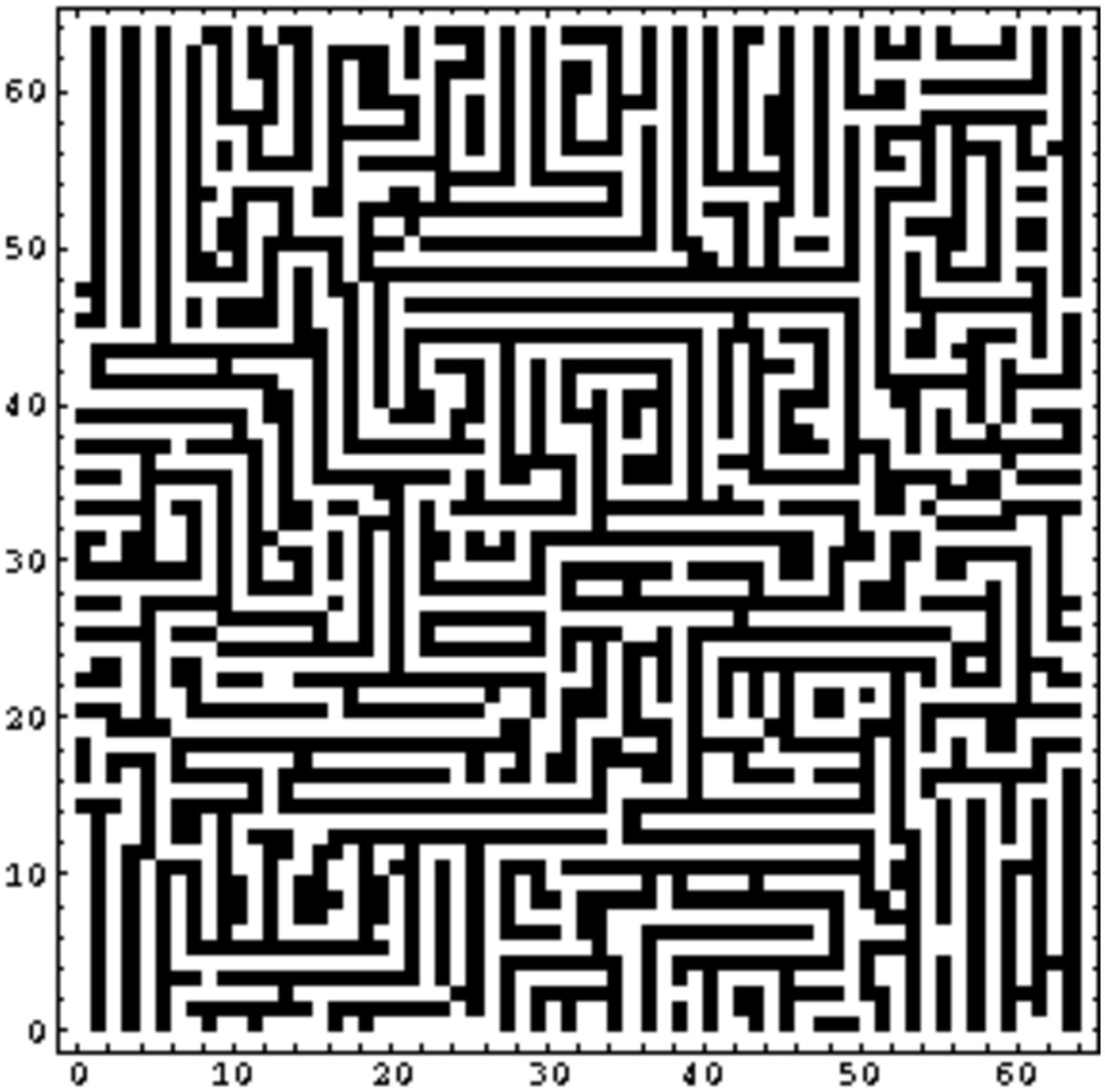}\hspace{0.5cm}\includegraphics[width=6cm,angle=0]{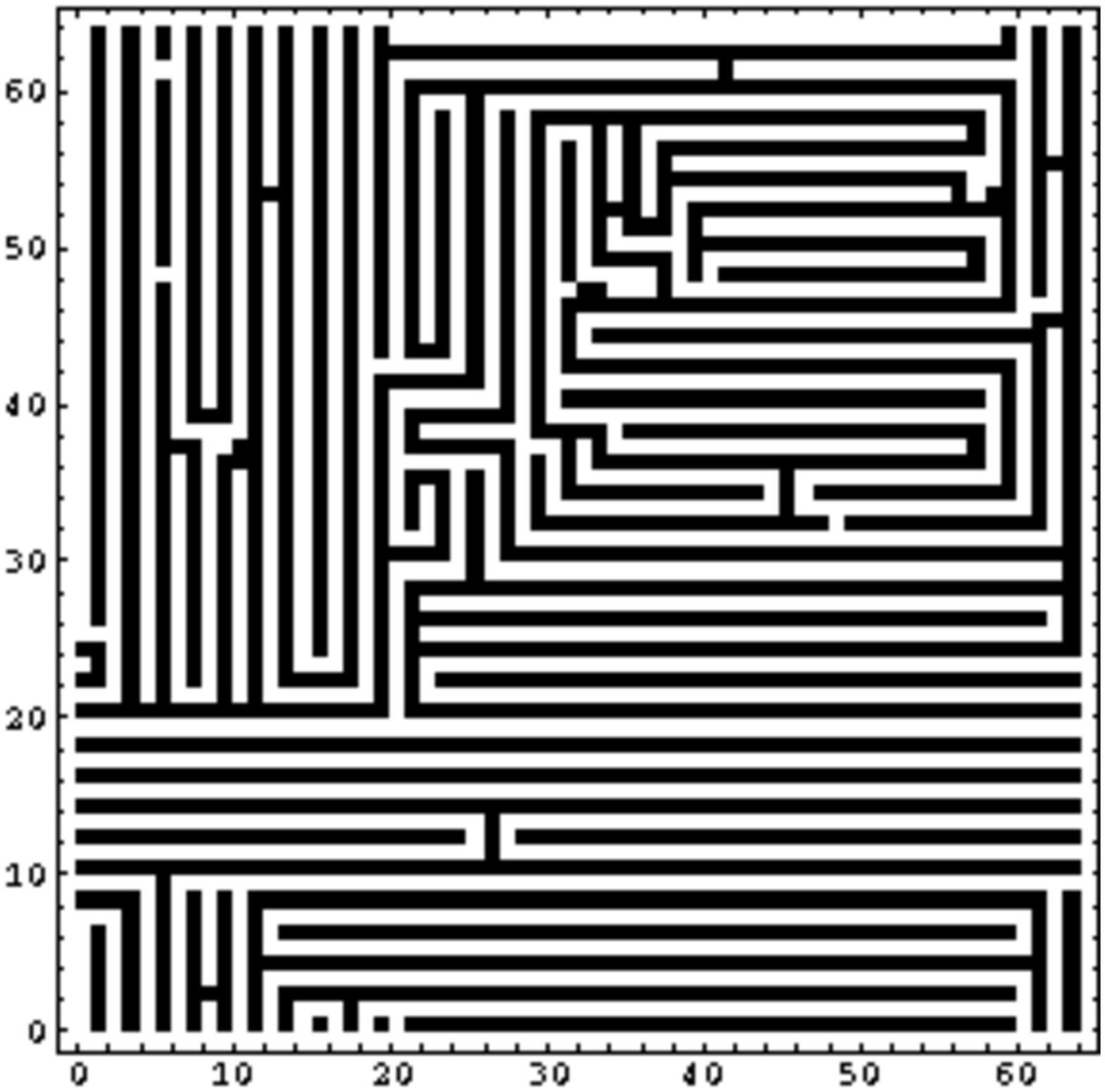}
\includegraphics[width=6cm,angle=0]{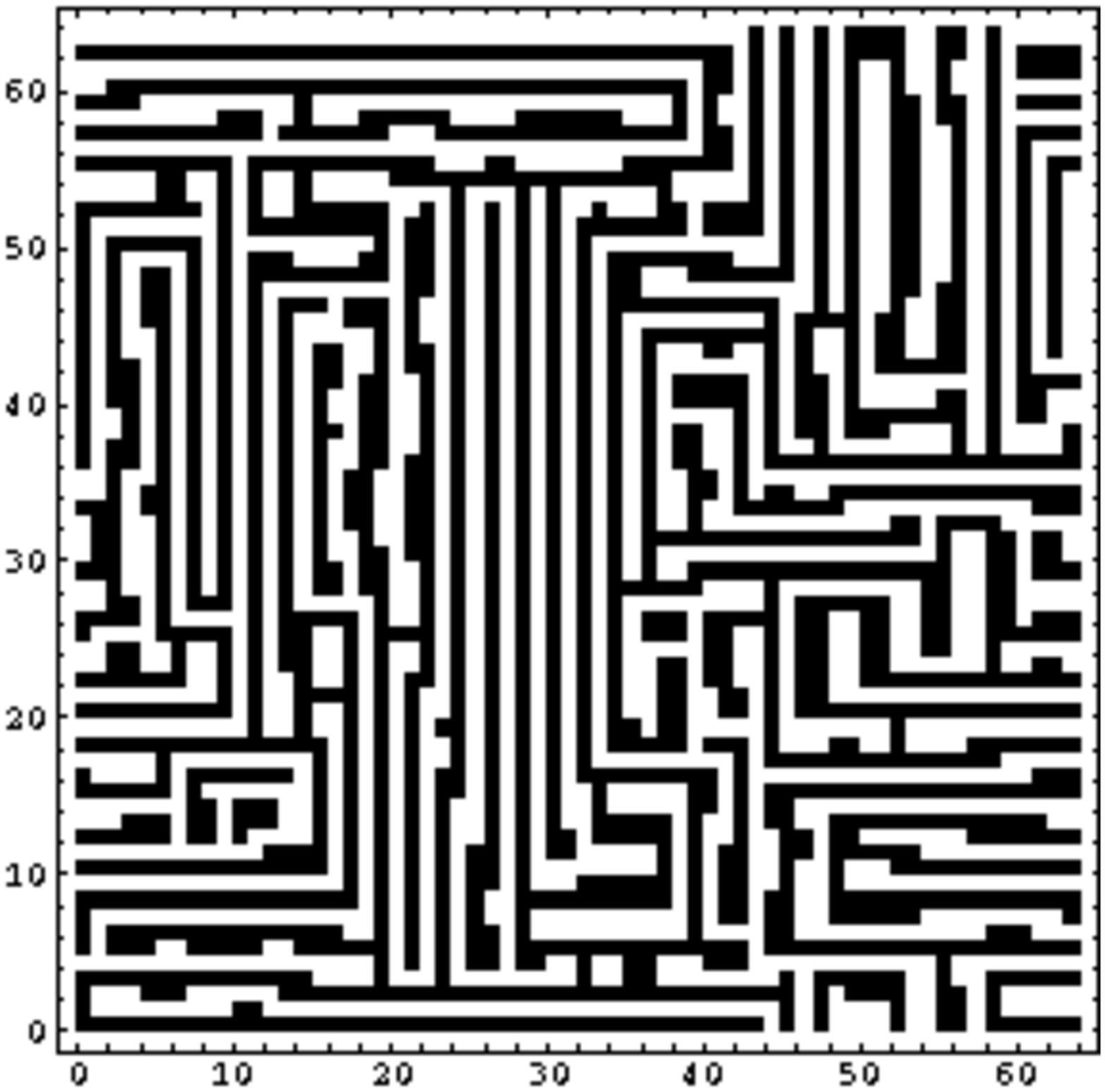}\hspace{0.5cm}\includegraphics[width=6cm,angle=0]{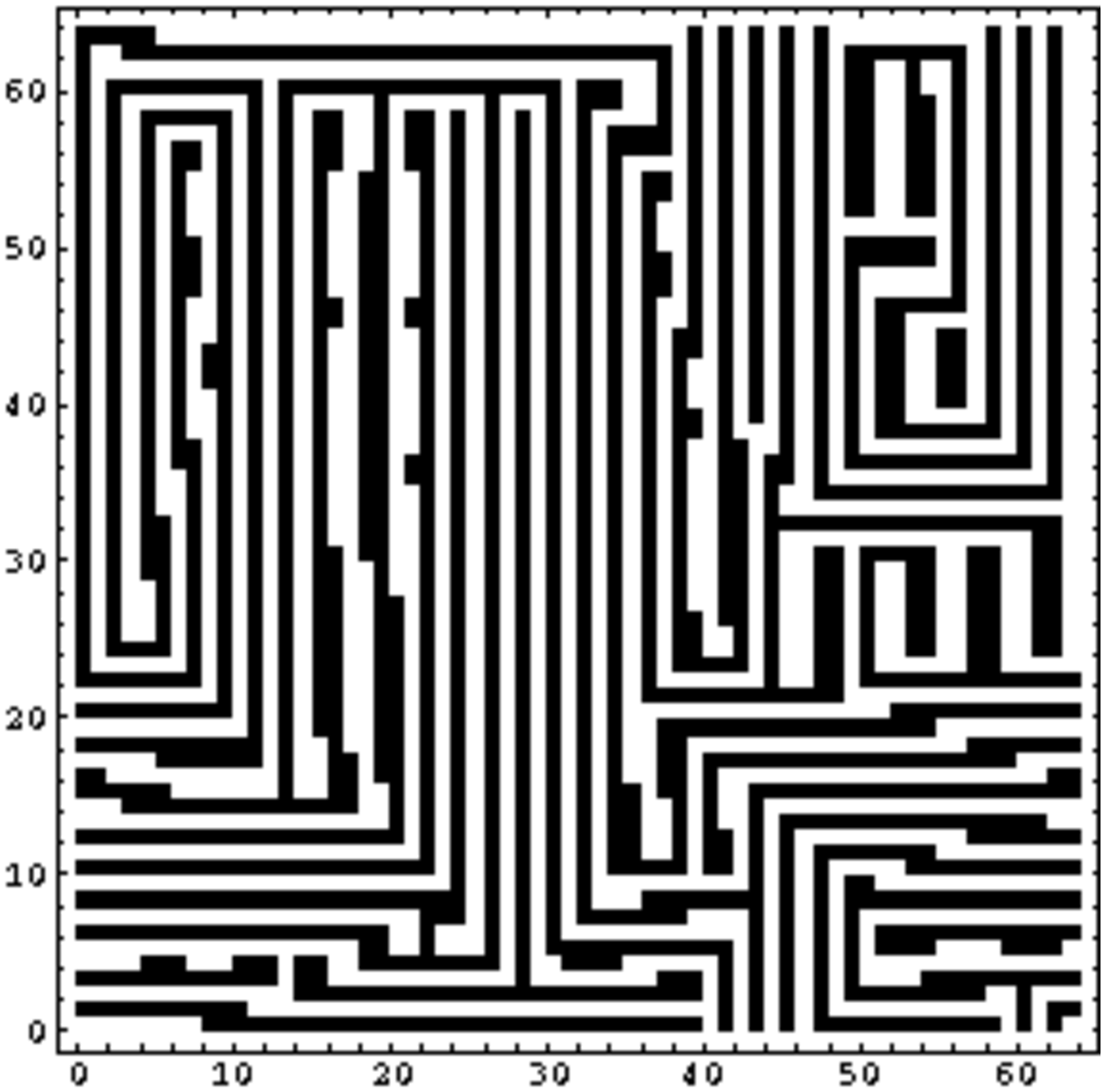}
\caption{Snapshot of a $64\times 64$ cooled at a constant rate
$r=0.0001$. Upper figures: $\delta=1.0$; lower figures
$\delta=1.25$. The left figures were obtained at $T=0.2$ and the
right ones at $T=0.1$.} \label{cooling-fig4}
\end{figure}

In order to understand in more details which is the mechanism
responsible for this drastic slowing down, let us take a direct
look into the configurations. In Fig. \ref{cooling-fig4} we see
four typical realizations of the spin configurations for the same
system of $N=64\times 64$ spins with $\delta=1$ and $\delta=1.25$,
respectively. The snapshots presented were obtained at $T=0.2$ and
$T=0.1$. What we observe is that in the region of metastability
($\delta = 1.25$), domains of $h=2$ structures develop which
prevent the system to reach a larger order, being responsible of
slowing down. On the other hand, we see that in the $\delta=1.0$
case, these structures rapidly disappear, during the fast
relaxation regime at intermediate temperature.

Concluding, we can see that the slow cooling experiments confirm
that the metastability is the responsible of the slow relaxation
dynamics observed in this two-dimensional model.

\section{Discussion and further directions}

In this work we have presented a rather detailed description of
the present state of the art concerning the macroscopic properties
of the two dimensional Ising model with competition between
nearest--neighbors interactions and long--range antiferromagnetic
interactions in a square lattice. Our analysis focused on the
nature of the different thermodynamical transitions for low values
of $\delta$ and on the out of equilibrium dynamics in the low
temperature phases of the model.

One of the most interesting thermodynamical properties is the weak
first order nature of the order-disorder phase transition for some
range of values of $\delta$, predicted by extensive Monte Carlo
numerical simulations. This result appears to answer the apparent
contradiction between the prediction of  Abanov \cite{Abanov}  and
Booth {\em et al.} results \cite{Booth}. The challenge persist to
determine experimentally the nature of this transition in
ultrathin magnetic films \cite{Vaterlaus}. A question that remains
open refers to the order of the phase transition for large values
of $\delta$. While a continuos approximation predicts that the
transition is first order for any value of $\delta$ on very
general grounds \cite{Cannas1}, the Monte Carlo results appears to
be consistent with a change in the order of the transition at some
finite value of $\delta$, above which the transition becomes
continuos. Anyway, if the transition for large values of $\delta$
is so weak that it becomes indistinguishable from a continuos one,
the difference could be irrelevant from the experimental point of
view.

The existence of a first order transition of this type is also
interesting for other reason. This behavior strongly resembles
that observed in the three dimensional Coulomb frustrated
ferromagnet \cite{Grousson,Viot}. Hence, it opens the possibility
to observe in a two dimensional model dynamical phenomena similar
to that seen in fragile glassforming liquids \cite{Grousson2}.
Works along this line are in progress.

Another interesting point about the thermodynamics of this model is
related to the existence of metastable striped states at low
temperature for large values of $\delta$, which generate complex
dynamical behaviors. As we have seen, the existence of metastable
states at low temperatures reveals the true nature of the different
dynamical regimes observed inside the striped phases. Some early
works seemed to show contradictory evidence on the dynamical
behavior of the system. On one side \cite{Toloza}, there was evidence that for
small values of $\delta$ the system presents a behavior similar to
that observed in glassy systems (logarithmic growth of the linear
domain size). On the other side, in \cite{Stariolo} the authors
found a behavior proper of a simple coarsening  process (algebraic
domain growth) in the same region. We have shown how different
types of dynamical analysis allows one to interpret these
apparently contradictory results, showing that the differences
observed can be actually ascribed to the existence, at low
temperatures, of a metastability region where metastable stripe
states of different width coexist. The analysis of the striped
domains growth shows that these metastable states trap the
dynamics of the system during certain transient that diverges as
$T\to 0$. When the system is quenched to the metastable region,
small domains of the metastable phase pin the domain walls of the
stable phase; such metastable domains generate free energy
barriers whose height is independent of the domain size $L$, thus
slowing the coarsening process for a finite
(temperature-dependent) period of time. Measurements taken with
observation time scales smaller than this characteristic periods
display an apparently logarithmic behavior. Numerical experiments
in which a slow cooling is performed from a thermalized state
above the transition temperature give further support to this
interpretation (see Fig.\ref{cooling-fig4}).

The existence of multiple  metastable ordered states at higher
values of $\delta$ and very low temperatures opens other
interesting perspectives. We expect that the domain
walls of the stable phases should be pinned by clusters of the
different coexisting metastable phases. Therefore, we expect
a slow dynamics characterized by multiple characteristic time
scales, associated with the different free energy barriers
generated by each metastable phase.

We see that many of the  open questions  remarked above appear for
large values of $\delta$. Unfortunately, in this region finite
size effects and the strong dynamical slowing down make numerical
simulations prohibitive, since larger system sizes should be
considered. The usage of more sophisticated non-Metropolis
sampling methods, like a recent adaptation of Creutz cluster
algorithm \cite{Stoycheva} or the multicanonical algorithm
\cite{Berg}, may help, at least for studying equilibrium
properties.

 Finally, one important point of experimental importance \cite{Portmann}
 that remains to be answered concerns
 the role of defects in the high temperature order-disorder
 transition. Some advances along this line has been done
 recently in in the triangular lattice \cite{Stoycheva}.

This work was partially supported by grants from Consejo Nacional
de Investigaciones Cient\'{\i}ficas y T\'ecnicas CONICET
(Argentina), Agencia C\'ordoba Ciencia (C\'ordoba, Argentina)  and
Secretar\'{\i}a de Investigaci\'on Cient\'{\i}fica y Tecnol\'ogica
de la Universidad Nacional de C\'ordoba (Argentina). P.M. Gleiser
thanks Fundaci\'on Antorchas (Argentina) for financial support. We
wish to thank H. Toloza, D. Stariolo and M. Montemurro for their
fruitful contributions.


\begin{thebibliography}{99}
\bibitem{differences} It is worth noting that the definition  of the Hamiltonian in
\cite{DeBell,Macisaac,Booth} is slightly different from
ours. While in those papers the dipolar term contains a sum over
all pairs of spins, here
 we consider the sum over every pair of spins just once. This
leads to the equivalence $\delta = J/2$, $J$ being the exchange
parameter in the above references. Since the dipolar parameter
also fixes the energy units in our work, there is also a factor
$1/2$ between the critical temperatures obtained in those works
and ours.
\end{thebibliography}
\end{document}